\documentclass[aps,10pt,prd,twocolumn,nofootinbib,notitlepage,superscriptaddress]{revtex4-2}

\usepackage{graphicx}
\usepackage{dcolumn}
\usepackage{bm}
\usepackage{amsmath,amssymb}
\usepackage{slashed}
\usepackage{xcolor}
\usepackage{physics}
\usepackage{multirow}
\usepackage[colorlinks=true, pdfstartview=FitV, bookmarks=true, bookmarksnumbered=true, breaklinks]{hyperref}
\usepackage{mathtools,braket}
\usepackage{soul}
\usepackage{lipsum}  
\usepackage{color}
\definecolor{blue}{rgb}{0.0, 0.0, 1.0}
\definecolor{red}{rgb}{1.0, 0.0, 0.0}
\definecolor{royalblue}{rgb}{0.0, 0.14, 0.4}

\usepackage{hyperref}
\hypersetup{colorlinks=true,citecolor=blue,linkcolor=blue,urlcolor=blue}

\usepackage[mathlines]{lineno}
\usepackage{tikz,xcolor,hyperref}
\usepackage{mathrsfs}
\usepackage{tikz}
\definecolor{lime}{HTML}{A6CE39}
\DeclareRobustCommand{\orcidicon}{%
	\begin{tikzpicture}
	\draw[lime, fill=lime] (0,0) 
	circle [radius=0.16] 
	node[white] {{\fontfamily{qag}\selectfont \tiny ID}};
	\draw[white, fill=white] (-0.0625,0.095) 
	circle [radius=0.007];
	\end{tikzpicture}
	\hspace{-2mm}
}
\foreach \x in {A, ..., Z}{%
	\expandafter\xdef\csname orcid\x\endcsname{\noexpand\href{https://orcid.org/\csname orcidauthor\x\endcsname}{\noexpand\orcidicon}}
}

\begin{document}
%========================================================================================
\title{Electromagnetic structure of charged and neutral strange vector mesons}
%========================================================================================

%----------------------------------------------------------------------------------------
\author{Parada~T.~P.~Hutauruk\orcidA{}} 
\email[E-mail:]{phutauruk@hiroshima-u.ac.jp}
\affiliation{International Institute for Sustainability with Knotted Chiral Meta Matter (WPI-SKCM$^2$), Hiroshima University, Higashi-Hiroshima, Hiroshima 739-8526, Japan}
%========================================================================================

\date{\today}

%========================================================================================
\begin{abstract} 
We investigate the electromagnetic form factors of the charged $K^{*+}$(892) (u$\bar{s}$) and neutral $K^{*0}$(896) (d$\bar{s})$ strange vector mesons within the covariant Nambu–Jona-Lasinio model, employing the Schwinger proper-time regularization scheme to regularize ultraviolet divergences that incorporated the QCD aspect of the quark confinement. To this end, we evaluate the charge (electric) $G_C(Q^2)$, magnetic moment $G_M(Q^2)$, and quadrupole moment $G_Q(Q^2)$ form factors and charge radii of the charged $K^{*+}$(892) and neutral $K^{*0}$(896) strange vector mesons. We find that the charge radii are $\big<r^2 \big>_{K^{*+}} =$ 0.45 fm$^2$ and $\big< r^2 \big>_{K^{*0}} =$ -0.04 fm$^2$, respectively, while the magnetic moments are found $\mu_{K^{*+}} =$ 2.67 $\mu_N$ and $\mu_{K^{*0}} =$ 0.032 $\mu_N$, consistent with other theoretical calculations and lattice QCD simulation results.
\end{abstract}
\maketitle

%========================================================================================
\section{Introduction} \label{sec:intro}
%========================================================================================
Studies of the internal structure of hadrons in terms of quark and gluon degrees of freedom, which is one of the playgrounds of quantum chromodynamics (QCD), have a long history and attract the interest of the hadron community. Key theoretical tools used to probe hadron structure, namely, parton distribution functions (PDFs)~\cite{Lorce:2025aqp,Hutauruk:2016sug,Collins:1981uw,Hutauruk:2025wkn}, electromagnetic form factors (EMFFs)~\cite{Hernandez-Pinto:2024kwg,Hutauruk:2025bjd}, fragmentation functions (FFs)~\cite{Aschenauer:2019kzf,Metz:2016swz,Echevarria:2016scs}, generalized parton distributions~\cite{Sun:2020jng,Shi:2023oll,Diehl:2015uka,Son:2024uet}, transverse-momentum-dependent parton distributions~\cite{Ninomiya:2017ggn,Liu:2025fuf,Boussarie:2023izj,Bacchetta:2019sam,Echevarria:2016scs}, and generalized transverse-momentum-dependent distributions~\cite{Bertone:2025vgy,Linek:2024dzs,Hatta:2024vzv}. These observables provide us with crucial information on parton dynamics in QCD bound states and also offer deep insights into nonperturbative QCD at low energies~\cite{Ioffe:2005ym,Gross:2022hyw,Leutwyler:2012ax}. More recently, significant attention has focused on hadronic gravitational form factors (GFFs) of the spin-1 particle~\cite{Epelbaum:2021ahi,Pagels:1966zza,Cosyn:2019aio}, which encode information on mechanical properties such as mass, energy, angular momentum, and internal force distributions. In addition to GFFs, EMFFs provide complementary information on charge distributions. In that sense, they exhibit a richer spin structure than pseudoscalar mesons. The vector mesons with spin-1 are characterized by three form factors, after imposing current conservation, whereas pseudoscalar mesons are described by a single form factor.

The EMFFs of the $\rho^+$(770) vector meson have been extensively studied in both free space and nuclear medium using a variety of theoretical approaches~\cite{Xu:2024vkn,Bhagwat:2006pu,Cardarelli:1994yq,Brodsky:1992px,Hutauruk:2025bjd,Tanisha:2025qda,Hernandez-Pinto:2024kwg,DeMelo:2018bim,Krutov:2018mbu,Choi:2004ww,Hecht:1997uj,Braguta:2004kx,Aliev:2004uj,Jaus:2002sv,Carrillo-Serrano:2015uca,Hutauruk:2025wkn,Cloet:2014rja,Shi:2023oll,Roberts:2011wy}. In contrast, relatively few studies~\cite{Hawes:1998bz,Bhagwat:2006pu} have addressed the EMFFs of the $K^{*+}$(892) strange vector mesons. The $K^{*+}$(892) strange mesons differ from the $\rho^+$(770) meson in their quark flavor composition, although both are spin-1 states. In particular, the $K^{*+}$(892) meson is composed of a light quark and a strange antiquark ($u\bar{s}$), leading to distinct quark-gluon dynamics that influence its internal structure and electromagnetic properties. This is largely due to the heavier strange-quark mass, which introduces significant flavor symmetry breaking effects in the form factors. Similar to the $\rho^+$(770) meson, the $K^{*+}$(892) and $K^{*0}$(896) strange mesons are characterized by three EMFFs, corresponding to the charge, magnetic dipole moment, and quadrupole moment. Experimentally, however, extracting the $K^{*+}$(892) and $K^{*0}$(896) strange meson EMFFs remains challenging due to their short lifetimes, which partly explains the limited attention they receive compared to that for the pseudoscalar mesons. So, at present, they remain largely unexplored and are not yet rigorously determined, in particular the EMFFs of the $K^{*0}$(896) meson.

From the experimental perspective, data on the EMFFs of the $K^{*+}$(892) and $K^{*0}$(896) strange vector mesons are very limited compared to that for the pseudoscalar mesons~\cite{Amendolia:1984nz,NA7:1986vav,JeffersonLabFpi-2:2006ysh}. Precise measurements of the EMFFs for the strange vector mesons are expected from future facilities, including COMPASS/AMBER++ at CERN~\cite{Adams:2018pwt}, the Jefferson Lab 12 GeV program (with a possible 22 GeV upgrade)~\cite{Accardi:2023chb}, J-PARC~\cite{Sawada:2016mao}, the Electron-Ion Collider in China (EicC)~\cite{Anderle:2021wcy}, and the Electron-Ion Collider (EIC) at Brookhaven National Laboratory (BNL)~\cite{Arrington:2021biu}. In addition to upcoming experimental efforts, studies of the $K^{*+}$(892) and $K^{*0}$(896) strange meson EMFFs have also been carried out in lattice QCD simulations~\cite{Hedditch:2007ex,Detmold:2017oqb,Wang:2025hew,Meissner:2026zos,Luschevskaya:2026kxx,Lee:2008qf}. These developments will provide an important opportunity to confront theoretical predictions of this study with lattice QCD simulation results, thereby improving our understanding of the internal structure of the $K^{*+}$(892) and $K^{*0}$(896) strange vector mesons and, more broadly, the dynamics of QCD. Very recently, a comprehensive discussion on extracting the vector meson form factors on the lattice has been presented in Ref.~\cite{Meissner:2026zos}.

In this paper, we investigate the EMFFs of the $K^{*+}$(892) and $K^{*0}$(896) strange vector mesons, including the charge (electric), magnetic moment, and quadrupole moment form factors, as well as its charge radius, within the framework of the covariant Nambu–Jona-Lasinio (NJL) model, is also known as a chiral effective quark theory of QCD. In this work, the Schwinger proper-time regularization scheme is employed to regularize ultraviolet divergences that mimic the QCD aspects of quark confinement. The NJL model is a covariant framework that incorporates key nonperturbative features of QCD, most notably the spontaneous breaking of chiral symmetry. The model has been successfully applied to a wide range of hadronic and nuclear phenomena, including PDFs and charge symmetry breaking~\cite{Hutauruk:2018zfk}, FFs~\cite{Ninomiya:2017ggn}, nuclear matter and neutron star properties~\cite{Bentz:2001vc}, and phase transitions in dense matter~\cite{Whittenbury:2015ziz}. Using this framework, we compute the $K^{*+}$(892) and $K^{*0}$(896) strange meson EMFFs and charge radius. Furthermore, we compare our results with the lattice QCD simulation results and other theoretical predictions.

The structure of this paper is organized as follows. In Sec.~\ref{sec:vacuumNJL}, we briefly review the $K^{*+}$(892) and $K^{*0}$(896) strange vector mesons in the framework of the covariant NJL model, employing the Schwinger proper-time regularization scheme to regularize ultraviolet divergences and to simulate quark confinement. In Sec.~\ref{sec:ffv}, we present the general expressions for the EMFFs of the $K^{*+}$(892) and $K^{*0}$(896) strange vector mesons using the matrix element of the electromagnetic current, and we extract the charge (electric) $G_c(Q^2)$, magnetic moment $G_M(Q^2)$, and quadrupole moment $G_Q(Q^2)$ form factors and their charge radii $\big<r\big>$. In Sec.~\ref{sec:MR}, we provide our numerical results for the $K^{*+}$(892) and $K^{*0}$(896) strange vector meson EMFFs and other corresponding quantities. Finally, a summary and conclusion are given in Sec.~\ref{sec:summary}.

%========================================================================================
\section{SU(3) flavor NJL model} 
\label{sec:vacuumNJL}
%========================================================================================
We briefly describe the SU(3) flavor NJL model, starting from its effective Lagrangian. The NJL model is a Poincaré-covariant quantum field theory that incorporates key low-energy features of QCD. It explicitly captures dynamical chiral symmetry breaking (DCSB) properties of QCD that are manifested in the gap equation through the chiral condensate. This leads to the generation of constituent (dressed) quark masses from the current quark masses. The NJL model does not exhibit confinement. In this approach, confinement is effectively simulated through the Schwinger proper-time regularization scheme, which removes unphysical quark-production thresholds associated with hadron decay into free quarks.

The SU(3) flavor NJL effective Lagrangian is expressed in terms of local four-fermion contact interactions (gluonic degrees of freedom are integrated out and absorbed into effective coupling constants) given by
\begin{eqnarray}
    \label{eq:vacNJL1}
    \mathscr{L}_{\mathrm{NJL}} &=& \bar{\psi}_q \big( i \partial\!\!\!/ -\hat{m}_q \big) \psi_q \nonumber \\
    &+& G_\pi \big[ \big(\bar{\psi}_q \lambda_i \psi_q \big)^2 -\big( \bar{\psi}_q \gamma_5 \lambda_i \psi_q \big)^2 \big] \nonumber \\
    &-& G_\rho \big[ \big( \bar{\psi}_q \gamma^\mu \lambda_i \psi_q \big)^2 + \big( \bar{\psi}_q \gamma^\mu \gamma_5 \lambda_i \psi_q \big)^2 \big],
\end{eqnarray}
where $\psi^{T} = (\psi_u, \psi_d, \psi_s)$ denotes the quark field in flavor space with $q=u,d,s$, $\lambda_i$ are the Gell-Mann matrices ($i=1,\dots,8$) with $\lambda_0=\sqrt{2/3},\mathbf{1}$, and $\hat{m}q=\mathrm{diag}(m_u,m_d,m_s)$ is the current-quark mass matrix. Note that the four-fermion interaction term is proportional to the coupling constant $G_\pi$, which governs quark–antiquark interactions in the scalar and pseudoscalar channels and is responsible for dynamical chiral symmetry breaking (DCSB). The coupling $G_\rho$ corresponds to vector interaction channels.

Using the Hartree mean-field approximation, we evaluate the dressed quark masses through the quark self-energy. The corresponding gap equation in the Schwinger proper-time regularization scheme can be written as
\begin{eqnarray}
    \label{eq:vacuumNJL2}
    M_q &=& m_q + \frac{N_c G_\pi M_q}{\pi^2} \int_{\tau_{\mathrm{UV}}}^{\tau_{\mathrm{IR}}} \frac{d\tau}{\tau^2} e^{-\tau M_q^2},
\end{eqnarray}
where $\tau_{\mathrm{UV}} = \Lambda_{\mathrm{UV}}^{-2}$ and $\tau_{\mathrm{IR}} = \Lambda_{\mathrm{IR}}^{-2}$ denote the ultraviolet (UV) and infrared (IR) integration limits, respectively. The IR cutoff is fixed at $\Lambda_{\mathrm{IR}} = 0.24~\text{GeV}$, corresponding to $\Lambda_{\mathrm{QCD}}$, while $\Lambda_{\mathrm{UV}}$ is determined by fitting the pion mass $m_\pi = 140~\text{MeV}$ and decay constant $f_\pi = 93~\text{MeV}$. Note that $\Lambda_{\mathrm{UV}}$ and $\Lambda_{\mathrm{IR}}$ serve as regularization scales in the NJL model and are treated as intrinsic model parameters. The standard dressed quark propagator for a quark flavor $q=(l,s)$ is defined by
\begin{eqnarray}
    \label{eq:gap1}
    S_l (k) &=& \frac{[k\!\!\!/ - M_l]}{[k^2 - M_l^2 + i \epsilon]},  \\
    S_s (k) &=& \frac{[k\!\!\!/ - M_s]}{[k^2 - M_s^2 + i \epsilon]},
\end{eqnarray}
where the subscripts $l=(u,d)$ and $s$ denote the light and strange quark sectors, respectively.

The dressed quark–antiquark bound state of the $K^{*+}$(892) and $K^{*0}$(896) strange vector mesons can be solved using the Bethe–Salpeter equation (BSE) within the random phase approximation. In the vector-meson channel, the BSE solution corresponds to the two-body $t$-matrix of the interaction kernel. The resummation of the quark–antiquark bubble diagrams to all orders leads to a geometric series representation of the $t$-matrix, which can be expressed in reduced form,
\begin{eqnarray}
    \label{eq:vacuumNJL3}
    t_{K^*}^{\mu \nu} (p^2) &=& \Bigg[ \frac{-4iG_{\rho}}{1 + 2 G_{\rho} \Pi_{K^*} (p^2)} \Bigg] \nonumber \\
    &\times& \Bigg[ g^{\mu \nu} + 2 G_{\rho} \Pi_{K^*} (p^{2}) \frac{p^{\mu} p^{\nu}}{p^2}\Bigg], 
\end{eqnarray}
where $K^{*}$ stands for the $K^{*+}$(892) and $K^{*0}$(896) strange vector mesons. The polarization insertions (bubble diagrams) for the $K^*$ strange vector meson are given, respectively, by
\begin{eqnarray}
    \label{eq:vacuumNJL4}
      \Pi_{K^*} (p^2) P^{\mu \nu} \delta_{ab} &=& iN_c \int \frac{d^4k}{(2\pi)^4} \nonumber \\
      &\times& \mathrm{Tr} \Big[ \gamma^\mu \lambda_a S_{l} (p+k) \gamma^\nu \lambda_b S_{s} (k) \Big],
\end{eqnarray}
where $P^{\mu \nu} = g^{\mu \nu} - p^\mu p^\nu/p^2$ is the transverse projector, and $S_{l}(p+k)$ and $S_s(k)$ denote the dressed light and strange quark propagators, respectively, as defined in Eq.~(\ref{eq:gap1}). Note that the trace over color space has already been evaluated, while the remaining trace is taken over flavor and Dirac indices. Within the proper-time regularization scheme, the expression can be rewritten as
\begin{eqnarray}
    \Pi_{K^*} (p^2) &=& \frac{N_c}{2 \pi^2} \int_0^1 dx \int_{\tau_{\mathrm{UV}}}^{\tau_{\mathrm{Ir}}} \frac{d\tau}{\tau} \exp \big[ -\tau (C_1) \big],\nonumber \\
    &\times& \Big[ M_uM_s -(1-x)M_u^2 -xM_s^2 \nonumber \\
    &+&  2x(1-x) p^2 \Big],
\end{eqnarray}
where $C_1 =(xM_u^2 + (1-x) M_s^2-x(1-x)p^2)$. The $K^*$ strange meson mass can be determined straightforwardly from the pole position of the corresponding $t$-matrix in Eq.~(\ref{eq:vacuumNJL3}), which yields
\begin{eqnarray}
    \label{eq:vacuumNJL5}
    1 + 2 G_\rho \Pi_{K^*} \big( p^2 = m_{K^*}^2\big) &=& 0,
\end{eqnarray}
where the quantities are determined at the bound-state pole of the $t$-matrix. The mass of $m_{K^*}$ denotes the $K^*$ strange vector meson mass. To determine the $K^*$ strange vector meson–quark coupling constant, we expand the $t$-matrix in Eq.~(\ref{eq:vacuumNJL3}) around the pole at $p^2 = m_{K^*}^2$ (on-shell condition). The meson wave-function renormalization, equivalently the meson–quark coupling constant, is then obtained from
\begin{eqnarray}
    \label{eq:vacuumNJL8}
   Z_{K^*}^{-1} = \big[ g_{K^* qq}\big]^{-2} &=& - \frac{\partial \Pi_{K^* qq} (p^2)}{\partial p^2} \Bigg|_{p^2 = m_{K^*}^2},
\end{eqnarray}
where $g_{K^* qq}$ denotes the $K^*$ strange vector meson–quark coupling constant. Its expression in the proper time regularization scheme can be written as
\begin{eqnarray}
 \label{eq:vacuumNJL9}
    \big[g_{K^* qq}\big]^{-2} &=& \frac{N_c}{2 \pi^2} \int d\tau \int_0^1  dx x(1-x) \exp \big[-\tau (C_2) \big] \nonumber \\
    &\times& \Big[ M_u M_s -(1-x)Mu^2 -xM_s^2 \nonumber \\
    &-& 2x(1-x) m_{K^{*}}^2 +\frac{2}{\tau} \Big], 
\end{eqnarray}
with $C_2 = (xM_u^2 + (1-x) M_s^2 -x(1-x) m_{K^{*}}^2)$.

%========================================================================================
\section{Strange vector meson EMMFs}
\label{sec:ffv}
%========================================================================================
Here we present the EMFFs of the $K^*$ strange vector mesons in the NJL model. The general expression for the electromagnetic current of the $K^*$ strange vector mesons in the matrix element can be written in terms of three form factors,
\begin{eqnarray}
    \label{eq:vff1}
   \mathcal{J}_{K^*}^{\mu, \alpha \beta} (p'^{},p^{}) &=& \Big[ g^{\alpha \beta} F_{1K^*} (Q^2) -\frac{q^{\alpha } q^{\beta }}{2 m_{K^*}^{2}} F_{2K^*} (Q^2) \Big] \left( p'^{} + p^{}\right)^\mu \nonumber \\
   &-& \Big[ q^{\alpha } g^{\mu \beta} - q^{\beta } g^{\mu \alpha}\Big] F_{3 K^*} (Q^2),
\end{eqnarray}
where the Lorentz indices $\alpha$ and $\beta$ correspond to the polarization states of the incoming and outgoing $K^*$ strange vector meson, respectively, and $\mu$ denotes the photon polarization. The form factors $F_{1K^*}(Q^2)$, $F_{2K^*}(Q^2)$, and $F_{3K^*}(Q^2)$ characterize the $K^*$ strange vector meson internal structure. In the Sachs representation, these form factors can be combined into the charge (electric) form factor $G_C(Q^2)$, associated with the charge distribution, the magnetic form factor $G_M(Q^2)$, associated with the magnetization distribution, and the quadrupole form factor $G_Q(Q^2)$, which encodes information on the spatial deformation (shape) of the $K^*$ strange vector meson. They are defined, respectively, by
\begin{eqnarray}
    \label{eqvff2}
    G_Q^{}(Q^2) &=& F_{1K^*} (Q^2) + A F_{2K^*} (Q^2) -F_{3K^*} (Q^2), \\
    G_C^{} (Q^2) &=& F_{1K^*} (Q^2) + \frac{2}{3} \eta G_Q (Q^2), \\
    G_M^{} (Q^2) &=& F_{3K^*} (Q^2),
\end{eqnarray}
where $A = (1 + \eta)$ with $\eta = Q^{2}/(4 m_{K^*}^{2})$. It is worth noting that all form factors are dimensionless quantities.
\begin{figure}[ht]
\centering
\includegraphics[width=1\columnwidth]{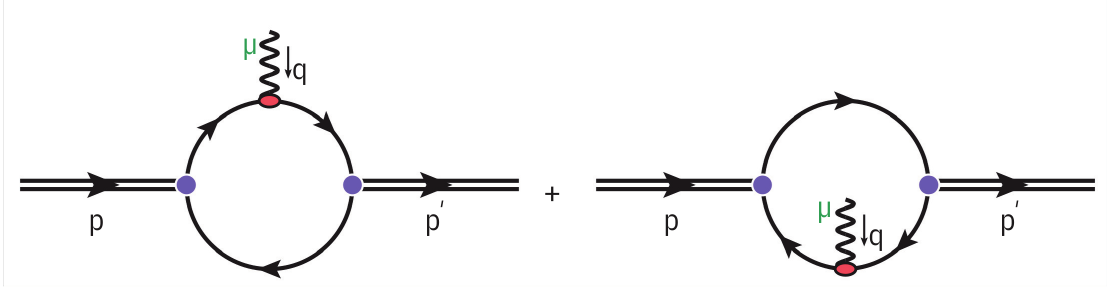} 
\caption{\label{fig6a} Two dominant Feynman diagrams for the electromagnetic current of the $K^*$ strange vector mesons. The purple shaded circles represent the Bethe–Salpeter vertices, while the red shaded oval is the quark–photon vertex.} 
\end{figure}

In the NJL model, and based on the two dominant Feynman diagrams shown in Fig.~\ref{fig6a}, the electromagnetic current of the $K^*$ strange vector mesons can be written as follows
\begin{eqnarray}
    \label{eqvff2a}
    \mathcal{J}^{\mu} (p',p) &=& i \int \frac{d^4k}{(2\pi)^4} \nonumber \\
    &\times& \mathrm{Tr} \Big[ \bar{\Gamma} S_{q_1}(p'+k) \Lambda^{\mu *}_{ Q} (p',p) \nonumber \\
    &\times& S_{q_1}(p +k) \Gamma S^{T}_{q_2}(-k) \Big],
\end{eqnarray}
where $\mathrm{Tr}$ denotes the trace over Dirac, color, and isospin indices, and the superscript $T$ indicates matrix transpose. The Bethe–Salpeter vertices of the $K^*$ strange vector meson are given by
\begin{eqnarray}
    \Gamma_{K^*}^\mu &=& g_{K{*}qq} \gamma^\mu \lambda_a,
\end{eqnarray}
while the $\Lambda^{\mu }_{\gamma Q} (p',p)$ stands for the dressed quark-photon vertex is defined as
\begin{eqnarray}
    \label{eqvff2b}
    \Lambda_i^{\big(\mathrm{BSE}\big) \mu } (Q^2) &=& \gamma^\mu F_{1i} (Q^2) + \frac{i\sigma^{\mu \nu} q_\nu}{2M} F_{2i} (Q^2), 
\end{eqnarray}
where the subscript $i=(\rho,\omega)$ denotes the dressed quark form factors obtained from the inhomogeneous Bethe–Salpeter equation. For a pointlike quark, one has $F_{1\omega}(Q^2)=1$ and $F_{2\rho}(Q^2)=F_{2\omega}(Q^2)=0$. This implies that $F_{2Q} (Q^2) = \frac{1}{6} F_{2\omega} (Q^2) \pm \frac{1}{2} F_{2 \rho} (Q^2)= 0$, where $Q = (U,D)$.

Finally, after a full derivation of the electromagnetic current for the $K^*$ strange vector meson in the NJL model, Eq.~(\ref{eqvff2a}), and comparison with the current in Eq.~(\ref{eq:vff1}), the $K^*$ strange vector meson form factors can be extracted. Taking into account the full structure of the dressed quark propagators, the compact expressions for the charged $K^{*+}$(892) strange vector meson form factors are given by
\begin{eqnarray}
F_{jK^{*+}} (Q^2) &=& \big[ F_{\mathrm{1U}} (Q^2) + F_{\mathrm{1S}} (Q^2)\big]  f_{jK^{*+}}^{V} (Q^2) \nonumber \\
&+& \big[ F_{\mathrm{2U}} (Q^2) + F_{\mathrm{2S}} (Q^2)\big] f_{jK^{*+}}^T (Q^2), 
\end{eqnarray}
where the index $j=(1,2,3)$. Analogously, for the neutral $K^{*0}$(896) strange vector meson, the form factor expression can be written as
\begin{eqnarray}
    F_{jK^{*0}} (Q^2) &=& \big[ F_{\mathrm{1D}} (Q^2) - F_{\mathrm{1S}} (Q^2)\big]  f_{jK^{*0}}^{V} (Q^2) \nonumber \\
&+& \big[ F_{\mathrm{2D}} (Q^2) - F_{\mathrm{2S}} (Q^2)\big] f_{jK^{*0}}^T (Q^2), 
\end{eqnarray}
where the vector body form factors for both $K^{*+}$(892) and $K^{*0}$(896) strange vector mesons in the Schwinger proper time regularization scheme are given by
\begin{eqnarray}
    \label{eq:bodyff}
    f_{1K^*}^{V} (Q^2) &=& - \frac{N_c g_{K* qq}^{2}}{8 \pi^2} \int_{\tau_{\mathrm{UV}}}^{\tau_{\mathrm{IR}}} d\tau \int_0^1 dx \int_{-x}^{x} dy \,\, A_{1V} \nonumber \\
    &\times& \exp\big[ -\tau (E_{1V}) \big] \nonumber \\
    &+& \frac{N_c g_{K^* qq}^{2}}{4\pi^2} \int_{\tau_{\mathrm{UV}}}^{\tau_{\mathrm{IR}}} \frac{d\tau}{\tau} \int_0^1 dx \nonumber \\
    &\times& \exp \big[ -\tau(E_{2V} )\big], 
\end{eqnarray}
\begin{eqnarray}
    f_{2K^*}^{V} (Q^2) &=& \frac{N_c g_{K^* qq}^{*2} m_{K^*}^{2}}{4\pi^2} \int_{\tau_{\mathrm{UV}}}^{\tau_{\mathrm{IR}}} d\tau \int_0^1 dx \int_{-x}^x dy \nonumber \\
    &\times& \big( x^2 -y^2 \big) \big( 1-x \big) \exp\Big[ -\tau \big( E_{1V} \big) \Big], 
\end{eqnarray}
\begin{eqnarray}
    f_{3K^*}^{V} (Q^2) &=& -\frac{N_c g_{K^* qq}^{2}}{8 \pi^2} \int^{\tau_{\mathrm{IR}}}_{\tau_{\mathrm{UV}}} d\tau \int_0^1 dx \int_{-x}^{x} dy \,\, A_{2V} \nonumber \\
    &\times& \exp\Big[ -\tau \big( E_1\big) \Big] \nonumber \\
    &+& \frac{3 N_c g_{K^* qq}^{2}}{4 \pi^2}  \int_0^1 dx \int_{\tau_{\mathrm{UV}}}^{\tau_{\mathrm{IR}}} \frac{d\tau}{\tau} \nonumber \\
    &\times& \exp\Big[ -\tau \big( E_{2V} \big)\Big],
\end{eqnarray}
where the variables $A_1$, $A_2$, $E_1$, and $E_2$ are defined as follow
\begin{eqnarray}
    A_{1V} &= & \Big[ \frac{2}{\tau} \big(1-x\big) - x m_{K^*}^{2} + (M_{s} -M_{l}) (2M_{s} \nonumber \\
    &-& x (M_{s} - M_{l})) \Big],  \\
    A_{2V} &=& \Big[ \frac{2}{\tau} (x+1)-x(1+2x) m_{K^*}^{*2} - (x^2 -y^2) \frac{Q^2}{2}  \\
    &+& (M_{s} - M_{l}) (x (M_{s} + M_{l}) + 2 M_{s} )\Big],  \\
    E_{1V} &=& \Big[ (x^2-x) m_{K*}^{*2} + \frac{1}{4} (x^2 -y^2) Q^2 + xM_{l}^{2} \nonumber \\
    &+& (1-x) M_{s}^{ 2}\Big],  \\
    E_{2V} &=& (x-x^2)Q^2 + M_{l}^{2},
\end{eqnarray}
 while the tensor body form factors are obtained by 
\begin{eqnarray}
    \label{eq:bodyffb}
    f_{1K^*}^{T} (Q^2) &=& - \frac{N_c g_{K^* qq}^{2} Q^2}{64 \pi^2 M_{q_1}} \int_0^1 dx \int_{-x}^{x} dy \int_{\tau_{\mathrm{UV}}}^{\tau_{\mathrm{IR}}} d\tau A_{1T}  \nonumber \\
    &\times& \exp \Big[ -\tau \big( E_{1V} \big)\Big],
\end{eqnarray}
\begin{eqnarray}
    f_{2K^*}^{T} (Q^2) &=& \frac{N_c g_{K^* qq}^{2}m_{K^*}^{2}}{16\pi^2 M_{l}} \int_0^1 dx \int_{-x}^{x} dy \int_{\tau_{\mathrm{UV}}}^{\tau_{\mathrm{IR}}} d\tau A_{2T} \nonumber \\
    &\times& \exp \Big[ -\tau \big( E_{1V} \big) \Big], 
\end{eqnarray}
\begin{eqnarray}
    f_{3 K^*}^{T} (Q^2) &=& - \frac{N_c g_{K^* qq}^{2}}{64 \pi^2 M_{l}} \int_0^1 dx \int_{-x}^{x} dy \int_{\tau_{\mathrm{UV}}}^{\tau_{\mathrm{IR}}} d\tau A_{3T} \nonumber \\
    &\times& \exp \Big[ -\tau \big( E_{1V}\big)\Big] \nonumber \\
    &+& \frac{N_c g_{K^* qq}^{2}}{16\pi^2M_{l}} \int_0^1 dx \int_{\tau_{\mathrm{UV}}}^{\tau_{\mathrm{IR}}} \frac{d\tau}{\tau} A_{4T} \nonumber \\
    &\times& \exp \Big[ -\tau \big( E_{2V} \big) \Big], 
\end{eqnarray}
where the variables $A_1T$, $A_{2T}$, $A_{3T}$, and $A_{4T}$ are defined, respectively, by
\begin{eqnarray}
    A_{1T} &=& \big((1-x)M_{s} + x M_{l}\big), \\
    A_{2T} &=& \big( M_{s} -xM_{l} + (x^2 -2x)(M_{s} - M_{l})\big), \\
    A_{3T} &=& \Big[ xM_{l} Q^2 - 2M_{s} m_{K^*}^{2} + 2(M_{s} - M_{l})\Big], \\
    A_{4T} &=& \big(M_{l} -(M_{s} -M_{l})\big).
\end{eqnarray}

It is worth noting that, in the strange-quark sector, the tensor contribution vanishes when the virtual photon couples to the strange quark, since $F_{2S}(Q^2)=0$. In other words, pion-cloud effects are not included in this work. The strange-quark tensor form factor $F_{2S}(Q^2)$ appears only when pion-cloud corrections are taken into account through the inhomogeneous Bethe–Salpeter equation. The inhomogeneous BSE-dressed quark form factors can be defined by
\begin{eqnarray}
\label{eq:vff4}
F_{1U} (Q^2) &=& \frac{1}{6} F_{1\omega} (Q^2) + \frac{1}{2} F_{1 \rho} (Q^2),  \\
F_{1D} (Q^2) &=& \frac{1}{6} F_{1\omega} (Q^2) - \frac{1}{2} F_{1 \rho} (Q^2), \\
F_{2U} (Q^2) &=& \frac{1}{6} F_{2\omega} (Q^2) + \frac{1}{2} F_{2 \rho} (Q^2),  \\
F_{2D} (Q^2) &=& \frac{1}{6} F_{2\omega} (Q^2) - \frac{1}{2} F_{2 \rho} (Q^2), \\
F_{1S} (Q^2) &=& e_s F_{1\phi} (Q^2),
\end{eqnarray}
where the dressed quark form factors are given by
$F_{1\phi} (Q^2) = 1/[1+ 2G_\rho \Pi_{\phi} (p^2)]$, and $F_{1 \rho (1\omega)} (Q^2) = 1/\big[1 + 2 G_{\rho (\omega)} \Pi_{\rho (\omega)} (p^{2}) \big]$, and $F_{2\rho}(Q^2)=F_{2\omega}(Q^2)=0$. The form factors $F_{1\rho}(Q^2)$ and $F_{1\omega}(Q^2)$ exhibit poles at $p^{2}=m_{\rho}^{2}$ and $p^{2}=m_{\omega}^{2}$, respectively. The strange-quark electric charge is $e_s=-1/3$. It is worth noting that the NJL Bethe–Salpeter kernel does not yield Pauli form factors because it lacks tensor–tensor four-fermion interaction channels.

Next, we straightforwardly compute the static electromagnetic properties of the $K^{*+}$(892) and $K^{*0}$(896) strange vector mesons, such as the magnetic moment $\mu_{K^*}$, the quadrupole moment $\mathcal{Q}_{K^*}$, and the mean-square charge radius $\langle r_C^{2} \rangle_{K^{*}}$. These quantities are respectively defined by
\begin{eqnarray}
\label{eq:magmon}
    \mu_{K^*} &=& G^{K^*}_M (Q^2 =0)\,\, \frac{M_N}{m_{K^*}}, \\
    \mathcal{Q}_{K^*} &=& \frac{G_{Q} (Q^2=0)}{m^{ 2}_{K^{*}}}, \\
    \big< r_C^{2} \big>_{K^{*}} &=& -6 \frac{\partial G_C (Q^2)}{\partial Q^2} \Bigg|_{Q^2 =0},
\end{eqnarray}
where $M_N=$ 934 MeV denotes the nucleon mass in free space. It is worth noting that the charge (conservation) normalization condition must be satisfied, $G_C^{}(Q^{2}=0)=1$ for the $K^{*+}$(892) and $\rho^+$(770) vector mesons, and for the $K^{*0}$(896) strange vector meson, $Q_c^{K^{*0}}(0) =$0. Here, we include the $\rho^+$(770) for comparison.

%========================================================================================
\section{Numerical Result and Discussion} \label{sec:MR}
%========================================================================================
Numerical results for the charged $K^{*+}$(892) and neutral $K^{*0}$(896) strange vector meson EMFFs, including the charge radius $\langle r_{K^*} \rangle$, the charge form factor $G_C(Q^2)$, the magnetic form factor $G_M(Q^2)$, and the quadrupole form factor $G_Q(Q^2)$ are presented in Figs.~\ref{fig1}-\ref{fig6}. The NJL model parameters used in this work are $G_\pi$, $G_\omega$, and $G_\rho$, as in Refs.~\cite{Gifari:2024ssz,Hutauruk:2021kej,Hutauruk:2016sug,Hutauruk:2018zfk}. The infrared regulator in the Schwinger proper-time scheme is fixed at $\Lambda_{\mathrm{IR}}=240~\text{MeV}$, of the order of $\Lambda_{\mathrm{QCD}}$. The dynamical light-quark mass is taken to be $M_l=400~\text{MeV}$. The remaining parameters are determined by fitting to the physical observables $m_\pi=140~\text{MeV}$, $m_K=495~\text{MeV}$, $M_N=934.36~\text{MeV}$, $m_\rho=770~\text{MeV}$, $m_{K^{*+}}= $ 892 MeV, $m_{K^{*0}}=$ 896 MeV, and the pion decay constant $f_\pi=93~\text{MeV}$. The resulting fit yields an ultraviolet cutoff $\Lambda_{\mathrm{UV}}=645~\text{MeV}$, pion coupling constant $G_\pi=19.04\times10^{-6}~\text{MeV}^{-2}$, $G_\rho = 11.04 \times 10^{-6}$ MeV$^{-2}$, and strange constituent quark mass $M_s=611~\text{MeV}$. The current up- and strange-quark masses are taken as $m_u=16~\text{MeV}$ and $m_s=356~\text{MeV}$, respectively, following Refs.~\cite{Gifari:2024ssz,Hutauruk:2021kej,Hutauruk:2016sug,Hutauruk:2018zfk}. Using these parameters as input to the formulation of the $K^{*}$ meson-quark coupling in Eq.(~\ref{eq:vacuumNJL9}), we obtain $g_{\bar{K}^{*0} qq} (g_{K^{*+} qq})  =$ 3.106 (3.132), respectively.
\begin{figure}[t]
	\centering
	\includegraphics[width=1.02\columnwidth]{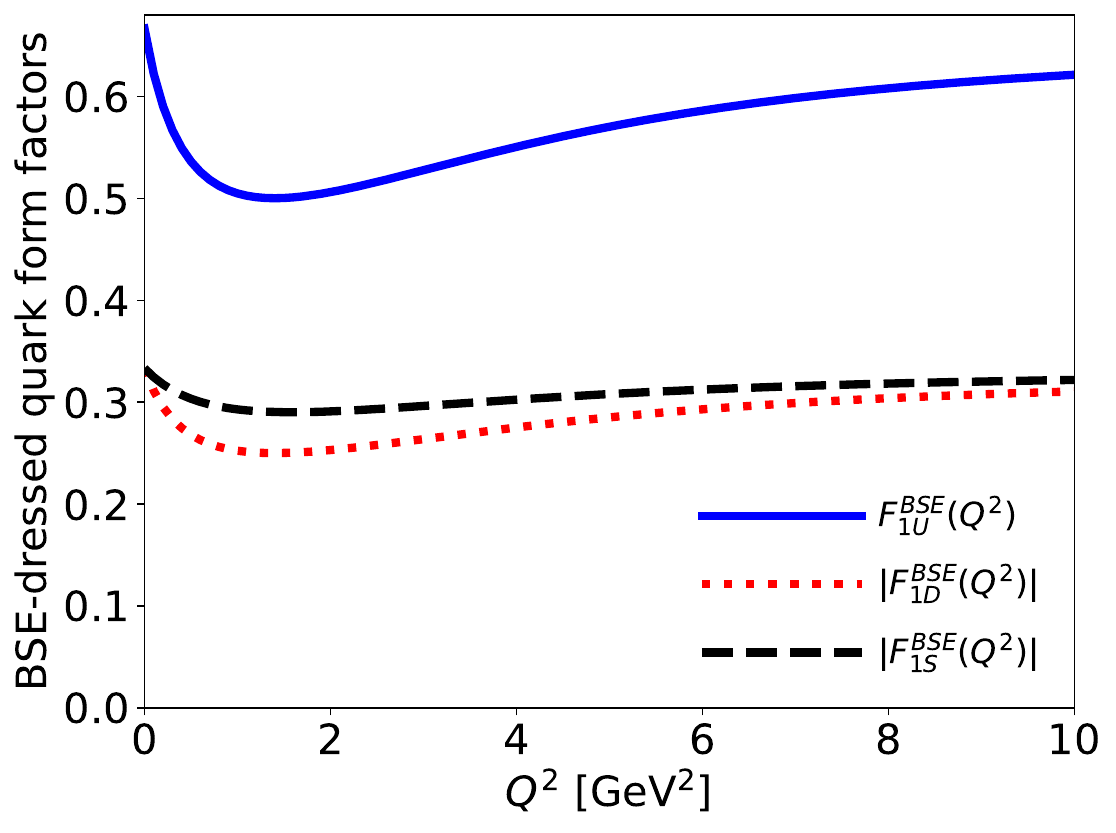}
 	\caption{\label{fig1a} Results for the BSE dressed quark form factors.}
\end{figure}

Results for the BSE dressed form factors of the up, down, and strange quarks are depicted in Fig.~\ref{fig1a}. It clearly shows that the $F_{1D}^{\rm{BSE}} (Q^2)$, $F_{1U}^{\rm{BSE}} (Q^2)$, and $F_{1S}^{\rm{BSE}} (Q^2)$ do not fall to zero at large $Q^2$, confirming the results obtained in Refs.~\cite{Hutauruk:2016sug,Cloet:2014rja,Carrillo-Serrano:2015uca,Hutauruk:2025bjd}. However, they behave as $F_{1D}^{\rm{BSE}} (Q^2) \simeq  e_d$, $F_{1U}^{\rm{BSE}} (Q^2) \simeq e_u$, and $F_{1S}^{\rm{BSE}} (Q^2) \simeq e_s$ at larger $Q^2$, where $e_d =-1/3$, $e_u = 2/3$, and $e_s = -1/3$ are the down, up, and strange quark charges, respectively. On the other hand, it indicates that the photon interacts with the bare (current) quark at larger $Q^2$. We find that these results are consistent with the expected results for the asymptotic freedom of QCD and those obtained in Refs.\cite{Hutauruk:2016sug,Cloet:2014rja,Carrillo-Serrano:2015uca,Hutauruk:2025bjd}.
\begin{figure}[t]
	\centering
	\includegraphics[width=1.02\columnwidth]{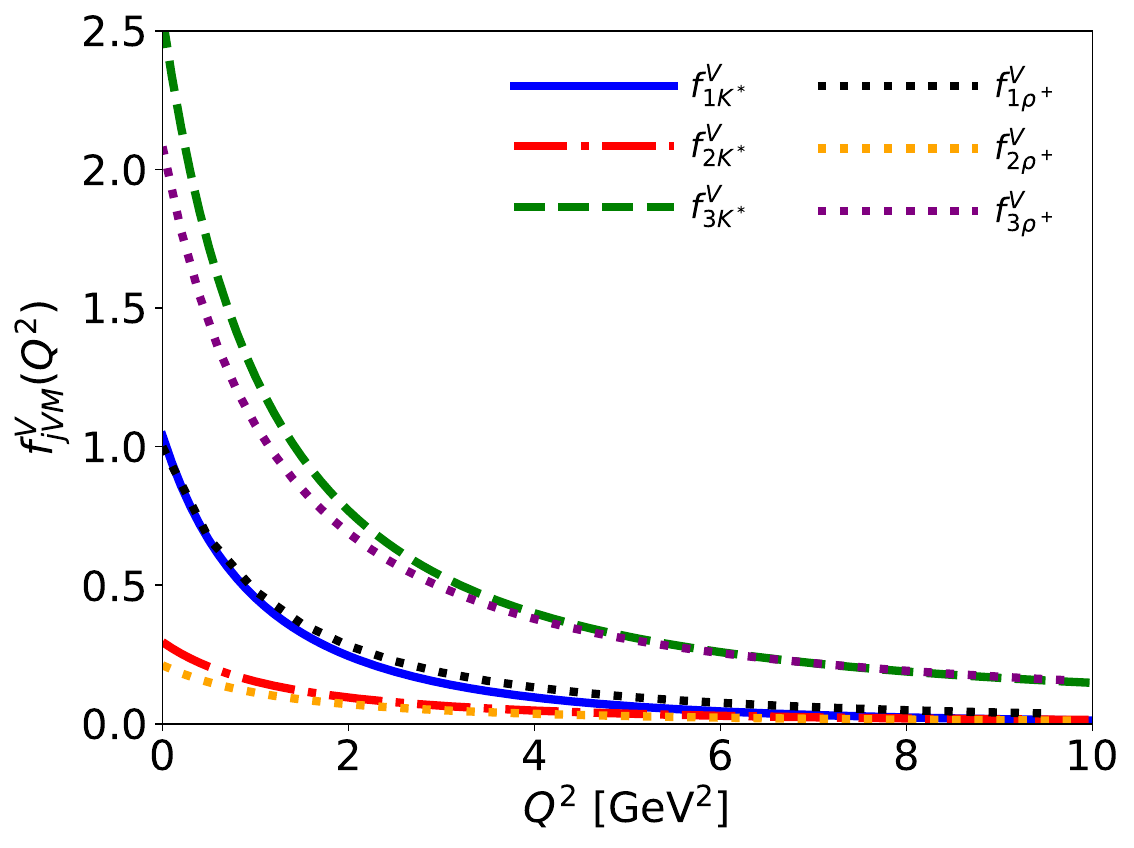}
 	\caption{\label{fig1} Results of the vector body form factors of the $K^{*}$ strange vector meson, which is relevant for the $K^{*0}$(896) and $K^{*+}$(892) vector mesons since the vector body for both is the same.}
\end{figure}

In Fig.~\ref{fig1}, we show the vector body form factors for the $K^{*0}$ (896) and $K^{*+}$(892) strange vector mesons. In addition, we also provide results for the $\rho^+$(770) vector meson for comparison. It is worth noting that the vector body form factors for the $K^{*0}$(896) strange vector meson are similar to those for the $K^{*+}$(892) vector meson. As shown in Fig.~\ref{fig1}, compared to the vector body form factors $f^V_{2\rho^{+}} (Q^2)$ and $f^V_{3\rho^{+}}(Q^2)$, the body form factors $f^V_{2K^*} (Q^2)$ and $f^V_{3K^*} (Q^2)$ have a larger value in the momentum transfer range $0 \leq Q^2 \leq 4$ GeV$^2$ and they have a similar values at $Q^2 > 4$ GeV$^2$. However, the vector body form factors of the $f^V_{1K^{*}} (Q^2)$ is found softer than that for the $f^V_{1\rho^{+}}(Q^2)$. Note that the vector body form factor satisfies charge conservation, resulting in $f^{V}_{1K^{*}} (Q^2) =$1, which is similar to the $\rho^{+}$ vector meson, giving the body form factor $f^{V}_{1\rho^{+}} (Q^2)=$ 1~\cite{Hutauruk:2025bjd,Cloet:2014rja,Carrillo-Serrano:2015uca}, while the vector body form factor $f_{3K^*}^V(0)=$ 2.55, which is a bit larger than that for the $f_{3\rho^{+}}^{V} (0) =$ 2.09. In addition, the body form factors $f_{3K^*}(0)$ and $f_{3\rho^{+}} (0)$ have slightly larger values in comparison to that for the canonical value of $\mu_1=2$ for a spin-1 particle, due to relativistic effects.
\begin{figure}[t]
	\centering
	\includegraphics[width=1.02\columnwidth]{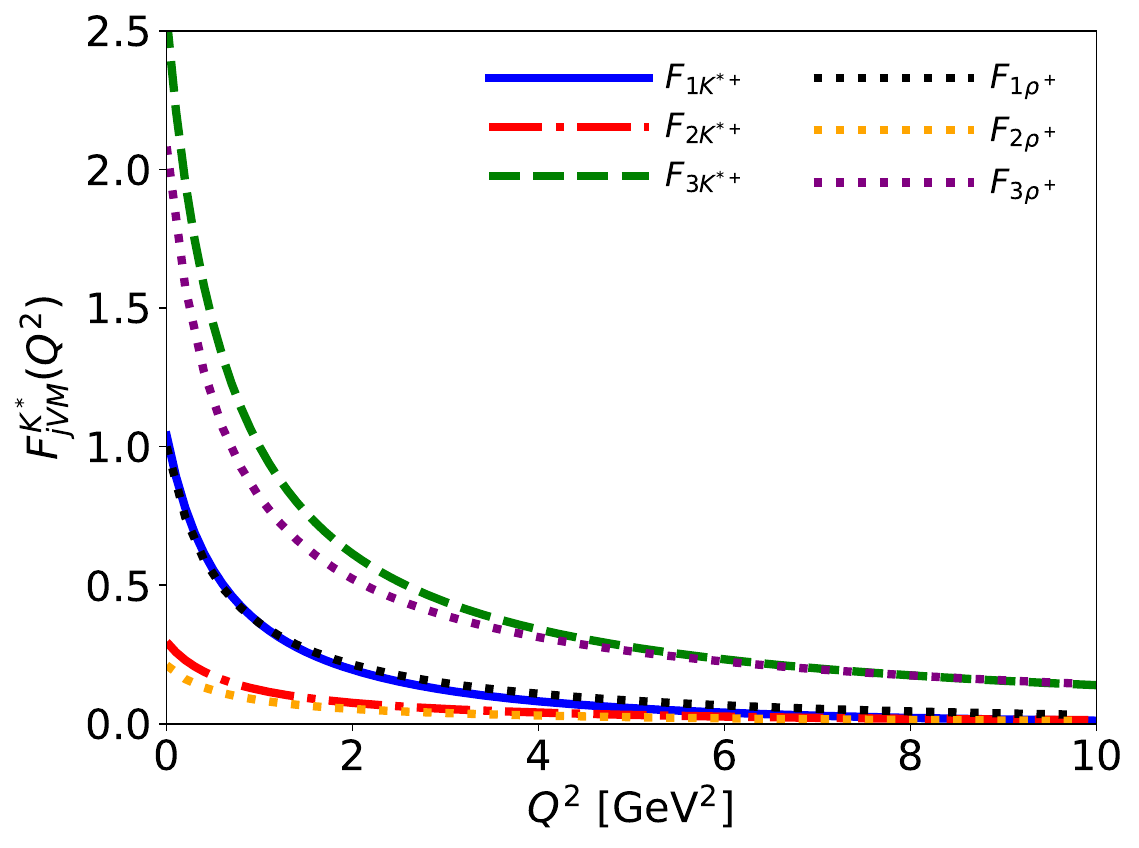}
 	\caption{\label{fig2}The BSE-dressed form factors for the $K^{*+}$(892) strange vector meson. Note the BSE-dressed form factors of the $\rho^{+}$(770) vector meson are shown for comparison.}
\end{figure}

The results of the BSE-dressed from factors for the $K^{*+}$(892) and $\rho^{+}$(770) vector mesons are depicted in Fig.~\ref{fig2}. Similar to the vector body form factor case, it shows that the value of the BSE-dressed form factor $F_{1K^{*+}} (Q^2)$ is a bit smaller than that for $F_{1\rho} (Q^2)$. However, again, both satisfy the normalization, giving $F_{1K^{*+}} (0) = F_{1\rho}(0) =$1. On the other hand, the dressed form factor $F_{2K^{*+}} (Q^2)$ is found to be larger than that for the $F_{2\rho^{+}}(Q^2)$ at around $0 \leq Q^2 \leq 4$ GeV$^2$. Such behavior is followed by the BSE-dressed form factor $F_{3K^{*}} (Q^2)$, resulting in larger values than those for the dressed form factor $F_{3\rho^{+}} (Q^2)$ in the corresponding $Q^2$ regime.
\begin{figure}[t]
	\centering
	\includegraphics[width=1.02\columnwidth]{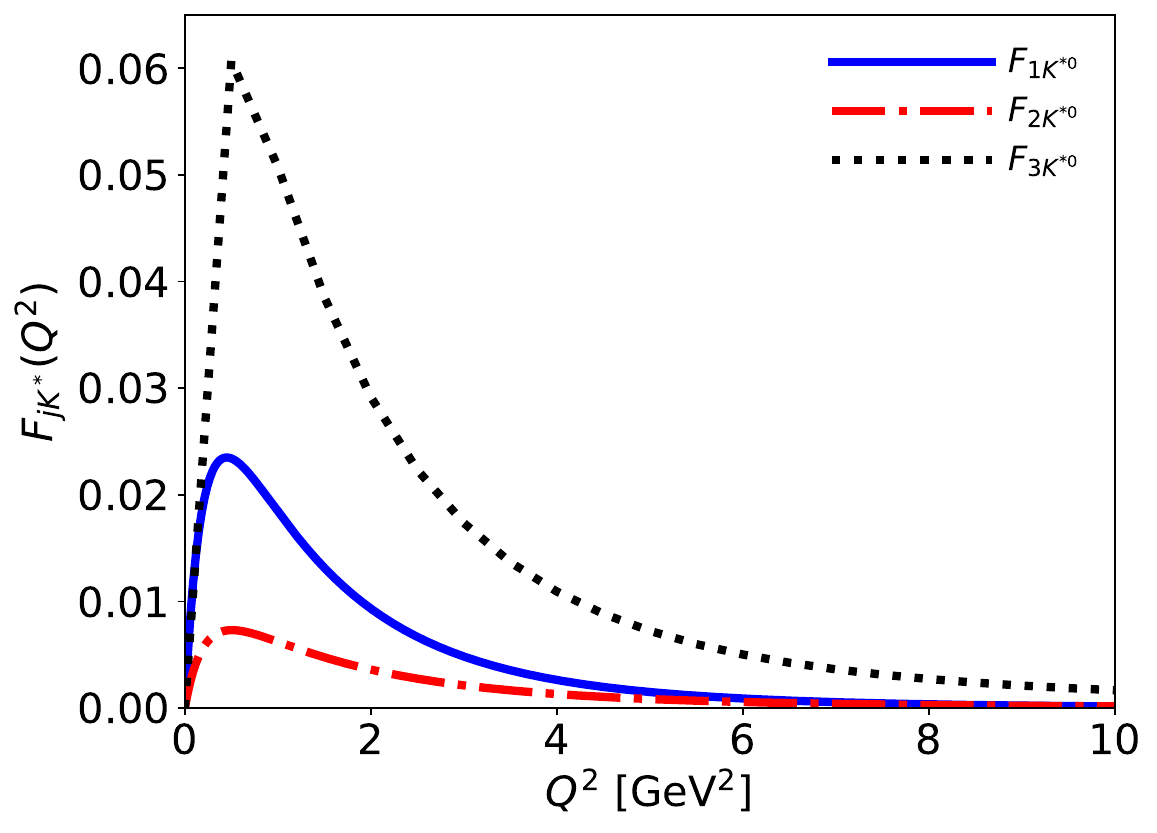}
 	\caption{\label{fig3} Same as in Fig.~\ref{fig2}, but only for the $K^{*0}$(896) strange vector meson.}
\end{figure}

Results for the BSE-dressed form factors of the $K^{*0} (896)$ strange vector meson are depicted in Fig.~\ref{fig3}. It certainly shows that the BSE-dressed form factors of the $K^{*0}$(896) strange vector meson are rather different from those for the $K^{*+}$(892) meson. For the $K^{*0}$(896) strange vector meson, it gives $F_{1K^{*0}} (0) = F_{2K^{*0}}(0) = F_{3K^{*0}} (0) =$0, as clearly shown in Fig.~\ref{fig3}. This is due to  $F_{1D}^{\rm{BSE}}(0)-F_{1S}^{\rm{BSE}}(0)$ is equal to zero. Moreover, we find that $F_{2K^{*0}} (Q^2)$ is the smallest among other dressed form factors, while $F_{3K^{*0}} (Q^2)$ has the largest value. 
\begin{figure}[ht]
	\centering
	\includegraphics[width=1.02\columnwidth]{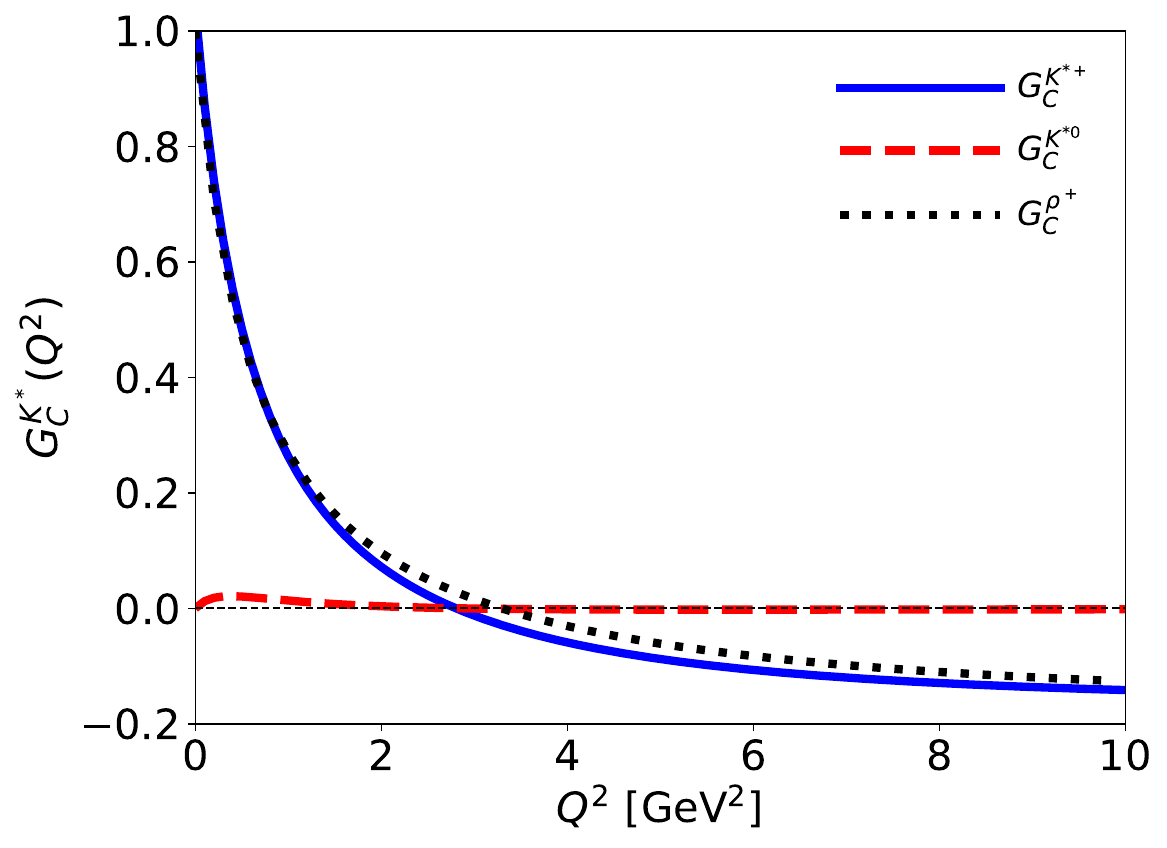}
 	\caption{\label{fig4} Results for the charge (electric) form factors $G_C (Q^2)$ for the $K^{*+}$(892), $\rho^+$(770), and $K^{*0}$(896) vector mesons. }
\end{figure}

Figure~\ref{fig4} shows the results of the charge (electric) form factors for the $K^{*+}$(892), $K^{*0}$(896), and $\rho^+$(770) vector mesons. Interesting features are shown by $G_{c}^{K^{*+}} (Q^2)$ in comparison with $G_{c}^{\rho^+} (Q^2)$ at around momentum transfer $Q^2 \simeq$ 3.0 GeV$^2$. It is found that the charge form factor $G_c^{K^{*+}}(Q^2)$ has a zero-crossing at around $Q^2 \simeq$ 3.2 GeV$^2$. Such a zero-crossing also occurred for the $\rho^+$(770) vector meson, but it decreases more slowly than that for the $K^{*+}$(892) vector meson. Also, we find that the charge form factor $G_c^{K^{*0}} (Q^2)$ shows different behavior in comparison to other charge form factors, where the nonzero charge form factor occurs at $0\,\, \mathrm{GeV}^2 <Q^2 <2$ GeV$^2$ for $G_c^{K^{*0}}(Q^2)$. It is worth noting that the charge from factors preserves the charge conservation by giving $G_c^{K^{*+}} (0) = G_c^{\rho^{+}} (0) =1$, while for the $K^{*0}$(896) vector meson, $G_{c}^{K^{*0}} (0) =$ 0. Similar behaviors of $G_C^{K^{*+}}(Q^2)$, $G_C^{K^{*0}}(Q^2)$, and $G_C^{\rho} (Q^2)$ are found in Refs.~\cite{Bhagwat:2006pu,Hawes:1998bz}. Using $G_C(0)$, the charge radii of $K^{*+}$(892) and $K^{*0}$ can be determined, and it gives $r_K^{*+} =$ 0.67 fm, which is consistent with that obtained in Ref.~\cite{Bhagwat:2006pu}, but lower than that obtained in Ref.~\cite{Luan:2015goa}. Moreover, our result of $r_{K^{*+}}$ is a bit higher than that obtained in Ref.~\cite{Hawes:1998bz,Gutierrez-Guerrero:2026rsb}. For $K^{*0}$(896) charge radius, we find that $\big<r^2\big>_{K^{*0}} = -0.04$ fm$^2$, which is consistent with other theoretical calculations in Refs.~\cite{Hawes:1998bz,Bhagwat:2006pu}.
\begin{figure}[t]
	\centering
	\includegraphics[width=1.02\columnwidth]{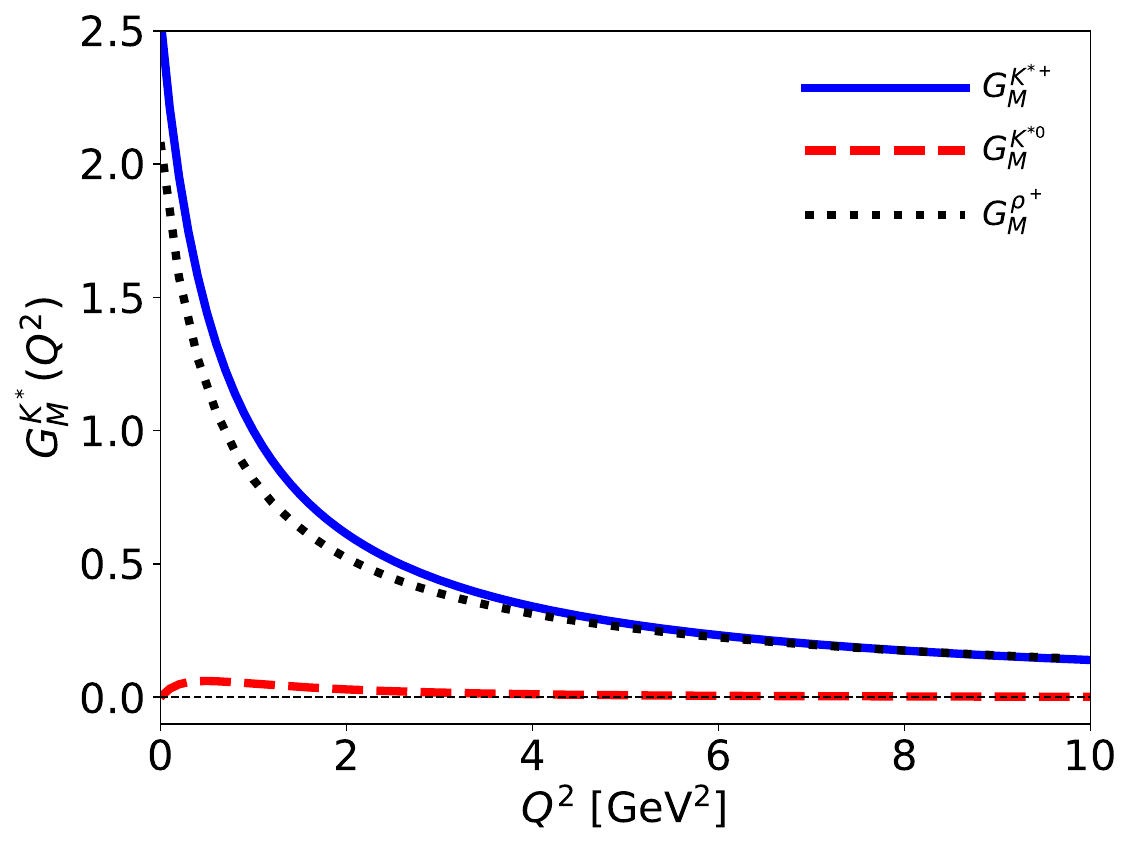}
 	\caption{\label{fig5} Results for the magnetic form factors $G_M (Q^2)$ for the $K^{*+}$(892), $\rho^{+}$(770), and $K^{*0}$(896) vector mesons.}
\end{figure}

In Fig.~\ref{fig5}, we show the results of the magnetic moment form factors for the $K^{*+}$(892), $\rho^+$(770), and $K^{*0}$(896) vector mesons. we find that the magnetic moment form factor $G_M^{K^{*+}} (Q^2)$ is larger than that for the $\rho^+$(770) meson at $0 <Q^2 < 4$ GeV$^2$, while for the $K^{*0}$(896) vector meson, the magnetic moment form factor $G_M^{K^{*0}}(Q^2)$ has the lowest values in comparison with others. This can be understood because the BSE-dressed form factors of the $K^{*0}$(896) and $K^{*+}$(892) strange vector mesons have very different magnitudes. Our results in $G_M^{K^{*}} (Q^2)$ are rather different from those obtained in Ref.~\cite{Bhagwat:2006pu}, having negative values of $G_M^{K^{*0}}(Q^2)$ at $Q^2 \leq 1$ GeV$^2$, while our result has positive value. For the result of $G_M^{K^{*+}} (Q^2)$, similar behaviors ar shown, decreasing as the $Q^2$ increasing. 
Note that the magnetic moment form factors of the $K^{*+}$(892) vector meson can be computed via Eq.~(\ref{eq:gap1}). We find the magnetic moment $\mu_{K^{*+}} =$ 2.67 $\mu_N$, while $\mu_{K^{*0}} =$ 0.032 $\mu_N$. Our results of $\mu_{K^{*+}}$ are comparable with those obtained in Refs.~\cite{Hawes:1998bz}, and consistent with the lattice result~\cite{Lee:2008qf}, while for $\mu_{K^{*0}}$, our result has comparable magnitude with the lattice simulation results~\cite{Hedditch:2007ex,Lee:2008qf}, but the lattice result has negative value, while our result has a positive value. Our result is rather different in magnitude from that obtained in Refs.~\cite{Hawes:1998bz,Bhagwat:2006pu,Aliev:2004uj,Badalian:2012ft}. A complete comparison of our results on $\mu_{K^{*+}}$ and $\mu_{K^{*0}}$ with other theoretical calculations and lattice results is provided in Table~\ref{tab1}.
\begin{figure}[t]
	\centering
	\includegraphics[width=1.02\columnwidth]{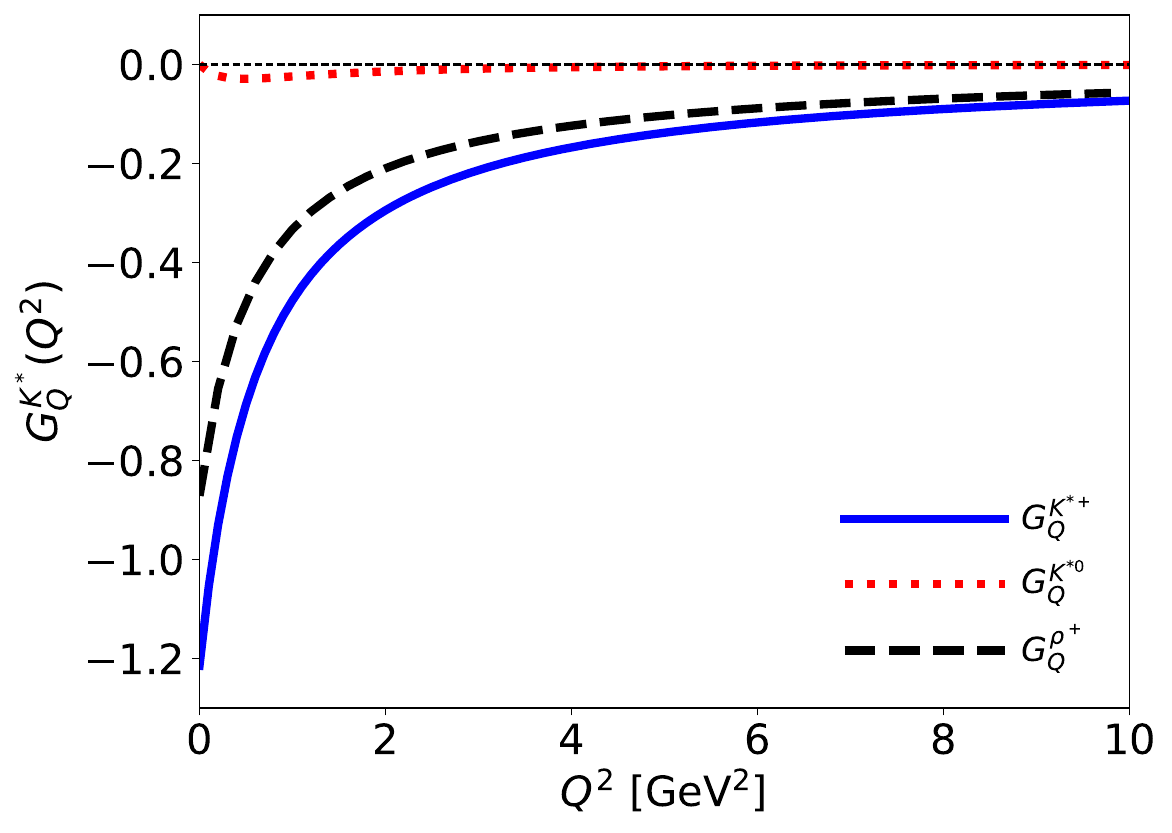}
 	\caption{\label{fig6} Results for the quadrupole moment form factors $G_Q (Q^2)$ for the $K^{*+}$(892), $\rho^+$(770), and $K^{*0}$(896) vector mesons.}
\end{figure}

In Fig.~\ref{fig6}, we show our results for the quadrupole moment form factors of the $K^{*+}$(892) and $K^{*0}$(896) strange vector mesons. We find the quadrupole moment form factor $G_Q^{\rho^{+}} (Q^2)$ has a larger value than that for $K^{*+}$(892) strange vector meson, $G_Q^{K^{*+}}(Q^2)$. For the $K^{*0}$(896) strange vector meson quadrupole moment form factor, it is found that the value of $G_Q^{K^{*0}}(Q^2)$ is the highest in comparison to others. We find that our results of the charged $K^{*+}$(892) and neutral $K^{*0}$(896) strange vector mesons are consistent with those obtained in Ref.~\cite{Bhagwat:2006pu}. We further find that our quadrupole moment $\mathcal{Q}_{K^{*+}}$ has good agreement with that obtained in Refs.~\cite{Hawes:1998bz,Bhagwat:2006pu,Gutierrez-Guerrero:2026rsb}, but we have higher values in comparison to others. For the $K^{*0}$(896) strange vector meson, it is found that the quadruple moment is consistent with that obtained in Ref.~\cite{Bhagwat:2006pu}, but rather different from that in Ref.~\cite{Hawes:1998bz}, as clearly shown in Table~\ref{tab1}.
\begin{table*}[t]
	\begin{ruledtabular}
		\renewcommand{\arraystretch}{1.5}
		\caption{Comparison of the present results on the magnetic moment $\mu_{K^*}$, quadrupole moment $\mathcal{Q}_{K^*}$, and charge radii $\big< r\big>_{K^*}$ with other theoretical and lattice QCD results. The units of $\mathcal{Q}_{K^{*}}$ and $\big< r \big>_{K^{*+}}$ are in fm$^2$ and fm, respectively. Note $\mu_N$ is the nuclear magneton}
		\label{tab1}
		\begin{tabular}{cc|ccc|ccc}
		  & \boldmath{Approaches} & $\mu_{K^{*+}} [\mu_N]$ & $\mathcal{Q}_{K^{*+}}$ & $r_{K^{*+}}$ & $\mu_{K^{*0}} [\mu_N]$ & $\mathcal{Q}_{K^{*0}}$ & $\big< r^2\big>_{K^{*0}}$ [$\rm{fm}^2$] \\[1ex] \hline \\[-1.5ex]
           This work & BSE-NJL model & 2.67 & $-0.06$ & 0.67  & 0.032 & -0.0007  & $-0.04$ \\
           Hawes \textit{et al.,}~\cite{Hawes:1998bz}  & DSE model (1998)  & 2.48 & $-0.03$ & 0.54 & $-0.42$ & 0.005  & $-0.05$ \\  
            Bhagwat \textit{et al.,}~\cite{Bhagwat:2006pu}  & DSE model (2008) & 2.34  & $-0.02$ & 0.66 & $-0.27$ & 0.0005 & $-0.08$ \\  
           Gutiérrez \textit{et al.,}~\cite{Gutierrez-Guerrero:2026rsb} & CI model (2026) & 2.28 & $-0.04$ & 0.54  & - &  - & - \\
           Luan \textit{et al.,}~\cite{Luan:2015goa} & Extended-NJL model (2015) & 2.37 & - & 1.05  & - &  - & - \\
           Lee~\cite{Lee:2008qf} & Lattice QCD (2008) & 2.81 (1) &- & -& $-0.07$& - & - \\
           Hedditch \textit{et al.,}~\cite{Hedditch:2007ex} & Lattice QCD (2007) & - &- &- & $-0.068(18)$ & - & - \\
           Aliev \textit{et al.,}~\cite{Aliev:2004uj} & LC-QCD Sum Rules & - & -& -& 0.28(4) &- & -\\
            Badalian \textit{et al.,}~\cite{Badalian:2012ft} & Relativistic Hamiltonian Approach & - & -& -& $-0.813$ &- & -
		\end{tabular}
	\end{ruledtabular}
\end{table*}

%========================================================================================
\section{Summary and Conclusion} \label{sec:summary}
%========================================================================================
To summarize, in this study, we have investigated the electromagnetic form factors (EMFFs) of the $K^{*+}$(892) and $K^{*0}$(896) strange vector mesons within the framework of the Nambu–Jona-Lasinio (NJL) model, employing the Schwinger proper-time regularization scheme to regularize ultraviolet divergences and to simulate QCD aspects of quark confinement. Using this framework, we evaluate the charge (electric) $G_c^{K^{*}}(Q^2)$, magnetic moment $G_M^{K^{*}}(Q^2)$, and quadrupole moment $G_Q^{K^{*}}(Q^2)$ form factors of the $K^{*+}$(892) and $K^{*0}$(896) strange vector mesons, as well as their charge radii.

We found that the charge (electric) form factor behaviors for $K^{*+}$(892), $\rho^{+}$(770), and $K^{*0}$(896) are similar to those found in Ref.~\cite{Hawes:1998bz,Bhagwat:2006pu}. We obtain charge radii $r_{K^{*+}} =$ 0.67 fm, consistent with calculations in Ref.~\cite{Bhagwat:2006pu}, but lower than that obtained in Ref.~\cite{Luan:2015goa}. It is also found that our results for $r_{K^{*+}}$ are a bit higher than those in Ref.~\cite{Hawes:1998bz,Gutierrez-Guerrero:2026rsb}. For
$K^{*0}$(896) charge radius, we find that $\big< r^2 \big>_{K^{*0}} = -0.04$ fm$^2$, which is consistent with other theoretical calculations in Refs.~\cite{Hawes:1998bz,Bhagwat:2006pu}.

For the magnetic moment, we found that $\mu_{K^{*+}} =$ 2.67 $\mu_N$, and $\mu_{K^{*0}} =$ 0.032 $\mu_N$, respectively. Our results for magnetic moment $\mu_{K^{*+}}$ are comparable with other theoretical calculations~\cite{Hawes:1998bz} and lattice QCD results~\cite{Lee:2008qf}, while for $\mu_{K^{*0}}$, it is found that our results are comparable in magnitude with the lattice results~\cite{Hedditch:2007ex,Lee:2008qf}.

The quadrupole moment form factors for $K^{*+}$(892) strange vector mesons are consistent with theoretical calculations~\cite{Gutierrez-Guerrero:2026rsb,Bhagwat:2006pu,Hawes:1998bz}, even though they are a bit different in magnitude. For the $K^{*0}$(896) strange vector meson, we found that our results are consistent with those obtained in Ref.~\cite{Bhagwat:2006pu}, but they have different signs.

The results of this study provide more information about the EMFFs of the $K^{*+}$(892) and $K^{*0}$(896) strange vector mesons, including the charge (electric) moment, magnetic moment, and quadrupole moment form factors. Since all these observables are not yet rigorously determined at present, more studies on the $K^{*0}$ strange vector meson using more sophisticated approaches and lattice QCD simulations are absolutely deserved.

%========================================================================================
\section*{Acknowledgements}
%========================================================================================
 This work was supported by the World Premier International Research Center Initiative (WPI) of Hiroshima University, MEXT, Japan. P.T.P.H. thanks the Asia Pacific Center for Theoretical Physics (APCTP) for its hospitality during a visit, where part of this work was carried out. 

%----------------------------------------------------------------------------------------
\bibliographystyle{elsarticle-num}
\bibliography{main}

@article{Bhagwat:2006pu,
    author = "Bhagwat, M. S. and Maris, P.",
    title = "{Vector meson form factors and their quark-mass dependence}",
    eprint = "nucl-th/0612069",
    archivePrefix = "arXiv",
    doi = "10.1103/PhysRevC.77.025203",
    journal = "Phys. Rev. C",
    volume = "77",
    pages = "025203",
    year = "2008"
}

@article{Tanisha:2025qda,
    author = "Tanisha and Puhan, Satyajit and Yadav, Anurag and Dahiya, Harleen",
    title = "{Valence quark distribution of light $\rho$ and heavy $J/\psi$ vector mesons in light-cone quark model}",
    eprint = "2505.09213",
    archivePrefix = "arXiv",
    primaryClass = "hep-ph",
    month = "5",
    year = "2025"
}

@article{Hutauruk:2016sug,
    author = "Hutauruk, Parada T. P. and Cloet, Ian C. and Thomas, Anthony W.",
    title = "{Flavor dependence of the pion and kaon form factors and parton distribution functions}",
    eprint = "1604.02853",
    archivePrefix = "arXiv",
    primaryClass = "nucl-th",
    reportNumber = "ADP-16-18-T973",
    doi = "10.1103/PhysRevC.94.035201",
    journal = "Phys. Rev. C",
    volume = "94",
    number = "3",
    pages = "035201",
    year = "2016"
}

@article{Hutauruk:2018zfk,
    author = {Hutauruk, Parada T. P. and Bentz, Wolfgang and Clo{\"e}t, Ian C. and Thomas, Anthony W.},
    title = "{Charge Symmetry Breaking Effects in Pion and Kaon Structure}",
    eprint = "1802.05511",
    archivePrefix = "arXiv",
    primaryClass = "nucl-th",
    reportNumber = "ADP-18-2-T1050",
    doi = "10.1103/PhysRevC.97.055210",
    journal = "Phys. Rev. C",
    volume = "97",
    number = "5",
    pages = "055210",
    year = "2018"
}

@article{Hutauruk:2021kej,
    author = "Hutauruk, Parada T. P. and Nam, Seung-il",
    title = "{Gluon and valence quark distributions for the pion and kaon in nuclear matter}",
    eprint = "2112.05435",
    archivePrefix = "arXiv",
    primaryClass = "hep-ph",
    doi = "10.1103/PhysRevD.105.034021",
    journal = "Phys. Rev. D",
    volume = "105",
    number = "3",
    pages = "034021",
    year = "2022"
}

@article{Gifari:2024ssz,
    author = "Gifari, Geoffry and Hutauruk, Parada T. P. and Mart, Terry",
    title = "{Nuclear medium meson structures from the Schwinger proper-time Nambu{\textendash}Jona-Lasinio model}",
    eprint = "2402.19048",
    archivePrefix = "arXiv",
    primaryClass = "hep-ph",
    doi = "10.1103/PhysRevD.110.014043",
    journal = "Phys. Rev. D",
    volume = "110",
    number = "1",
    pages = "014043",
    year = "2024"
}

@article{Xu:2024vkn,
    author = "Xu, Y. -Z. and Raya, K. and Rodr{\'\i}guez-Quintero, J. and Segovia, J.",
    title = "{Charge distributions of pseudoscalar and vector mesons from Dyson-Schwinger equations}",
    eprint = "2406.13306",
    archivePrefix = "arXiv",
    primaryClass = "hep-ph",
    doi = "10.1103/PhysRevD.110.054031",
    journal = "Phys. Rev. D",
    volume = "110",
    number = "5",
    pages = "054031",
    year = "2024"
}

@article{Cardarelli:1994yq,
    author = "Cardarelli, F. and Grach, I. L. and Narodetsky, I. M. and Salme, G. and Simula, S.",
    title = "{Electromagnetic form-factors of the rho meson in a light front constituent quark model}",
    eprint = "hep-ph/9502360",
    archivePrefix = "arXiv",
    reportNumber = "INFN-ISS-94-9A, INFN-ISS-94-9",
    doi = "10.1016/0370-2693(95)00230-I",
    journal = "Phys. Lett. B",
    volume = "349",
    pages = "393--399",
    year = "1995"
}

@article{Brodsky:1992px,
    author = "Brodsky, Stanley J. and Hiller, John R.",
    title = "{Universal properties of the electromagnetic interactions of spin one systems}",
    reportNumber = "SLAC-PUB-5763",
    doi = "10.1103/PhysRevD.46.2141",
    journal = "Phys. Rev. D",
    volume = "46",
    pages = "2141--2149",
    year = "1992"
}

@article{Hernandez-Pinto:2024kwg,
    author = "Hern{\'a}ndez-Pinto, R. J. and Guti{\'e}rrez-Guerrero, L. X. and Bedolla, M. A. and Bashir, A.",
    title = "{Electric, magnetic, and quadrupole form factors and charge radii of vector mesons: From light to heavy sectors in a contact interaction}",
    eprint = "2410.23813",
    archivePrefix = "arXiv",
    primaryClass = "hep-ph",
    doi = "10.1103/PhysRevD.110.114015",
    journal = "Phys. Rev. D",
    volume = "110",
    number = "11",
    pages = "114015",
    year = "2024"
}

@article{Gross:2022hyw,
    author = "Gross, Franz and others",
    title = "{50 Years of Quantum Chromodynamics}",
    eprint = "2212.11107",
    archivePrefix = "arXiv",
    primaryClass = "hep-ph",
    doi = "10.1140/epjc/s10052-023-11949-2",
    journal = "Eur. Phys. J. C",
    volume = "83",
    pages = "1125",
    year = "2023"
}

@article{Leutwyler:2012ax,
    author = "Leutwyler, H.",
    editor = "Zichichi, Antonino",
    title = "{On the history of the strong interaction}",
    eprint = "1211.6777",
    archivePrefix = "arXiv",
    primaryClass = "physics.hist-ph",
    doi = "10.1142/S0217732314300237",
    journal = "Mod. Phys. Lett. A",
    volume = "29",
    pages = "1430023",
    year = "2014"
}

@article{DeMelo:2018bim,
    author = "De Melo, J. P. B. C.",
    title = "{Unambiguous Extraction of the Electromagnetic Form Factors for Spin-1 Particles on the Light-Front}",
    eprint = "1810.11478",
    archivePrefix = "arXiv",
    primaryClass = "hep-ph",
    reportNumber = "LFTC-18-12/33",
    doi = "10.1016/j.physletb.2018.11.003",
    journal = "Phys. Lett. B",
    volume = "788",
    pages = "152--160",
    year = "2019"
}

@article{Krutov:2018mbu,
    author = "Krutov, A. F. and Polezhaev, R. G. and Troitsky, V. E.",
    title = "{Magnetic moment of the {\ensuremath{\rho}} meson in instant-form relativistic quantum mechanics}",
    eprint = "1801.01458",
    archivePrefix = "arXiv",
    primaryClass = "hep-ph",
    doi = "10.1103/PhysRevD.97.033007",
    journal = "Phys. Rev. D",
    volume = "97",
    number = "3",
    pages = "033007",
    year = "2018"
}

@article{Ioffe:2005ym,
    author = "Ioffe, B. L.",
    title = "{QCD at low energies}",
    eprint = "hep-ph/0502148",
    archivePrefix = "arXiv",
    doi = "10.1016/j.ppnp.2005.05.001",
    journal = "Prog. Part. Nucl. Phys.",
    volume = "56",
    pages = "232--277",
    year = "2006"
}

@article{Hawes:1998bz,
    author = "Hawes, F. T. and Pichowsky, M. A.",
    title = "{Electromagnetic form-factors of light vector mesons}",
    eprint = "nucl-th/9806025",
    archivePrefix = "arXiv",
    reportNumber = "ADP-98-24-T300, FSU-SCRI-98-054, FSU-SCRI-98-54",
    doi = "10.1103/PhysRevC.59.1743",
    journal = "Phys. Rev. C",
    volume = "59",
    pages = "1743--1750",
    year = "1999"
}

@inproceedings{Lorce:2025aqp,
    author = "Lorc{\'e}, C{\'e}dric and Metz, Andreas and Pasquini, Barbara and Schweitzer, Peter",
    title = "{Parton Distribution Functions and their Generalizations}",
    eprint = "2507.12664",
    archivePrefix = "arXiv",
    primaryClass = "hep-ph",
    month = "7",
    year = "2025"
}

@article{Collins:1981uw,
    author = "Collins, John C. and Soper, Davison E.",
    title = "{Parton Distribution and Decay Functions}",
    reportNumber = "OITS-166",
    doi = "10.1016/0550-3213(82)90021-9",
    journal = "Nucl. Phys. B",
    volume = "194",
    pages = "445--492",
    year = "1982"
}

@article{Sun:2020jng,
    author = "Sun, Baodong and Dong, Yubing",
    title = "{Generalized parton distribution functions of $\rho$ meson}",
    doi = "10.21468/SciPostPhysProc.3.014",
    journal = "SciPost Phys. Proc.",
    volume = "3",
    pages = "014",
    year = "2020"
}

@article{Shi:2023oll,
    author = "Shi, Chao and Li, Jicheng and Yin, Pei-Lin and Jia, Wenbao",
    title = "{Unpolarized generalized parton distributions of light and heavy vector mesons}",
    eprint = "2302.02388",
    archivePrefix = "arXiv",
    primaryClass = "hep-ph",
    doi = "10.1103/PhysRevD.107.074009",
    journal = "Phys. Rev. D",
    volume = "107",
    number = "7",
    pages = "074009",
    year = "2023"
}

@article{Ninomiya:2017ggn,
    author = {Ninomiya, Yu and Bentz, Wolfgang and Clo{\"e}t, Ian C.},
    title = "{Transverse-momentum-dependent quark distribution functions of spin-one targets: Formalism and covariant calculations}",
    eprint = "1707.03787",
    archivePrefix = "arXiv",
    primaryClass = "nucl-th",
    doi = "10.1103/PhysRevC.96.045206",
    journal = "Phys. Rev. C",
    volume = "96",
    number = "4",
    pages = "045206",
    year = "2017"
}

@article{Liu:2025fuf,
    author = "Liu, Wei-Yang and Zahed, Ismail",
    title = "{Tomography of the rho meson in the QCD instanton vacuum: Transverse momentum dependent parton distribution functions}",
    eprint = "2503.11959",
    archivePrefix = "arXiv",
    primaryClass = "hep-ph",
    doi = "10.1103/6ffp-qs8p",
    journal = "Phys. Rev. D",
    volume = "112",
    number = "3",
    pages = "034028",
    year = "2025"
}

@article{Boussarie:2023izj,
    author = "Boussarie, Renaud and others",
    title = "{TMD Handbook}",
    eprint = "2304.03302",
    archivePrefix = "arXiv",
    primaryClass = "hep-ph",
    reportNumber = "JLAB-THY-23-3780, LA-UR-21-20798, MIT-CTP/5386",
    month = "4",
    year = "2023"
}

@article{Aschenauer:2019kzf,
    author = "Aschenauer, Elke C. and Borsa, Ignacio and Sassot, Rodolfo and Van Hulse, Charlotte",
    title = "{Semi-inclusive Deep-Inelastic Scattering, Parton Distributions and Fragmentation Functions at a Future Electron-Ion Collider}",
    eprint = "1902.10663",
    archivePrefix = "arXiv",
    primaryClass = "hep-ph",
    doi = "10.1103/PhysRevD.99.094004",
    journal = "Phys. Rev. D",
    volume = "99",
    number = "9",
    pages = "094004",
    year = "2019"
}

@article{Metz:2016swz,
    author = "Metz, Andreas and Vossen, Anselm",
    title = "{Parton Fragmentation Functions}",
    eprint = "1607.02521",
    archivePrefix = "arXiv",
    primaryClass = "hep-ex",
    doi = "10.1016/j.ppnp.2016.08.003",
    journal = "Prog. Part. Nucl. Phys.",
    volume = "91",
    pages = "136--202",
    year = "2016"
}

@article{Bacchetta:2019sam,
    author = "Bacchetta, Alessandro and Bertone, Valerio and Bissolotti, Chiara and Bozzi, Giuseppe and Delcarro, Filippo and Piacenza, Fulvio and Radici, Marco",
    title = "{Transverse-momentum-dependent parton distributions up to N$^{3}$LL from Drell-Yan data}",
    eprint = "1912.07550",
    archivePrefix = "arXiv",
    primaryClass = "hep-ph",
    reportNumber = "JLAB-THY-19-3121",
    doi = "10.1007/JHEP07(2020)117",
    journal = "JHEP",
    volume = "07",
    pages = "117",
    year = "2020"
}

@article{Diehl:2015uka,
    author = "Diehl, Markus",
    title = "{Introduction to GPDs and TMDs}",
    eprint = "1512.01328",
    archivePrefix = "arXiv",
    primaryClass = "hep-ph",
    reportNumber = "DESY-15-234",
    doi = "10.1140/epja/i2016-16149-3",
    journal = "Eur. Phys. J. A",
    volume = "52",
    number = "6",
    pages = "149",
    year = "2016"
}

@article{Echevarria:2016scs,
    author = "Echevarria, Miguel G. and Scimemi, Ignazio and Vladimirov, Alexey",
    title = "{Unpolarized Transverse Momentum Dependent Parton Distribution and Fragmentation Functions at next-to-next-to-leading order}",
    eprint = "1604.07869",
    archivePrefix = "arXiv",
    primaryClass = "hep-ph",
    doi = "10.1007/JHEP09(2016)004",
    journal = "JHEP",
    volume = "09",
    pages = "004",
    year = "2016"
}

@article{Bertone:2025vgy,
    author = "Bertone, Valerio and Echevarria, Miguel G. and del Rio, {\'O}scar and Rodini, Simone",
    title = "{One-loop matching for leading-twist generalised transverse-momentum-dependent distributions}",
    eprint = "2502.07576",
    archivePrefix = "arXiv",
    primaryClass = "hep-ph",
    reportNumber = "IPARCOS-UCM-25-004, DESY-25-024",
    doi = "10.1007/JHEP05(2025)183",
    journal = "JHEP",
    volume = "05",
    pages = "183",
    year = "2025"
}

@article{Linek:2024dzs,
    author = {Linek, Barbara and {\L}uszczak, Marta and Sch{\"a}fer, Wolfgang and Szczurek, Antoni},
    title = "{Probing gluon GTMDs of the proton in deep inelastic diffractive dijet production at HERA}",
    eprint = "2403.15110",
    archivePrefix = "arXiv",
    primaryClass = "hep-ph",
    doi = "10.1103/PhysRevD.110.054027",
    journal = "Phys. Rev. D",
    volume = "110",
    number = "5",
    pages = "054027",
    year = "2024"
}

@article{Hatta:2024vzv,
    author = "Hatta, Yoshitaka and Yuan, Feng",
    title = "{Angular dependence in transverse momentum dependent diffractive parton distributions at small-x}",
    eprint = "2403.19609",
    archivePrefix = "arXiv",
    primaryClass = "hep-ph",
    doi = "10.1016/j.physletb.2024.138738",
    journal = "Phys. Lett. B",
    volume = "854",
    pages = "138738",
    year = "2024"
}

@article{Epelbaum:2021ahi,
    author = "Epelbaum, E. and Gegelia, J. and Mei{\ss}ner, U. -G. and Polyakov, M. V.",
    title = "{Chiral theory of {\ensuremath{\rho}}-meson gravitational form factors}",
    eprint = "2109.10826",
    archivePrefix = "arXiv",
    primaryClass = "hep-ph",
    doi = "10.1103/PhysRevD.105.016018",
    journal = "Phys. Rev. D",
    volume = "105",
    number = "1",
    pages = "016018",
    year = "2022"
}

@article{Pagels:1966zza,
    author = "Pagels, Heinz",
    title = "{Energy-Momentum Structure Form Factors of Particles}",
    doi = "10.1103/PhysRev.144.1250",
    journal = "Phys. Rev.",
    volume = "144",
    pages = "1250--1260",
    year = "1966"
}

@article{Cosyn:2019aio,
    author = "Cosyn, Wim and Cotogno, Sabrina and Freese, Adam and Lorc{\'e}, C{\'e}dric",
    title = "{The energy-momentum tensor of spin-1 hadrons: formalism}",
    eprint = "1903.00408",
    archivePrefix = "arXiv",
    primaryClass = "hep-ph",
    doi = "10.1140/epjc/s10052-019-6981-3",
    journal = "Eur. Phys. J. C",
    volume = "79",
    number = "6",
    pages = "476",
    year = "2019"
}

@article{Amendolia:1984nz,
    author = "Amendolia, S. R. and others",
    title = "{A Measurement of the Pion Charge Radius}",
    reportNumber = "CERN-EP/84-59",
    doi = "10.1016/0370-2693(84)90655-5",
    journal = "Phys. Lett. B",
    volume = "146",
    pages = "116--120",
    year = "1984"
}

@article{NA7:1986vav,
    author = "Amendolia, S. R. and others",
    editor = "Loken, S. C.",
    collaboration = "NA7",
    title = "{A Measurement of the Space - Like Pion Electromagnetic Form-Factor}",
    reportNumber = "CERN-EP-86-34",
    doi = "10.1016/0550-3213(86)90437-2",
    journal = "Nucl. Phys. B",
    volume = "277",
    pages = "168",
    year = "1986"
}

@article{JeffersonLabFpi-2:2006ysh,
    author = "Horn, T. and others",
    collaboration = "Jefferson Lab F(pi)-2",
    title = "{Determination of the Charged Pion Form Factor at Q**2 = 1.60 and 2.45-(GeV/c)**2}",
    eprint = "nucl-ex/0607005",
    archivePrefix = "arXiv",
    reportNumber = "JLAB-PHY-06-523",
    doi = "10.1103/PhysRevLett.97.192001",
    journal = "Phys. Rev. Lett.",
    volume = "97",
    pages = "192001",
    year = "2006"
}

@article{Accardi:2023chb,
    author = "Accardi, A. and others",
    title = "{Strong interaction physics at the luminosity frontier with 22 GeV electrons at Jefferson Lab}",
    eprint = "2306.09360",
    archivePrefix = "arXiv",
    primaryClass = "nucl-ex",
    reportNumber = "JLAB-PHY-23-3840, JLAB-THY-23-3848",
    doi = "10.1140/epja/s10050-024-01282-x",
    journal = "Eur. Phys. J. A",
    volume = "60",
    number = "9",
    pages = "173",
    year = "2024"
}

@article{Arrington:2021biu,
    author = "Arrington, J. and others",
    title = "{Revealing the structure of light pseudoscalar mesons at the electron{\textendash}ion collider}",
    eprint = "2102.11788",
    archivePrefix = "arXiv",
    primaryClass = "nucl-ex",
    doi = "10.1088/1361-6471/abf5c3",
    journal = "J. Phys. G",
    volume = "48",
    number = "7",
    pages = "075106",
    year = "2021"
}

@article{Anderle:2021wcy,
    author = "Anderle, Daniele P. and others",
    title = "{Electron-ion collider in China}",
    eprint = "2102.09222",
    archivePrefix = "arXiv",
    primaryClass = "nucl-ex",
    reportNumber = "Frontiers of Physics, Volume 16 Issue (6):64701, 2021",
    doi = "10.1007/s11467-021-1062-0",
    journal = "Front. Phys. (Beijing)",
    volume = "16",
    number = "6",
    pages = "64701",
    year = "2021"
}

@article{Sawada:2016mao,
    author = "Sawada, Takahiro and Chang, Wen-Chen and Kumano, Shunzo and Peng, Jen-Chieh and Sawada, Shinya and Tanaka, Kazuhiro",
    title = "{Accessing proton generalized parton distributions and pion distribution amplitudes with the exclusive pion-induced Drell-Yan process at J-PARC}",
    eprint = "1605.00364",
    archivePrefix = "arXiv",
    primaryClass = "nucl-ex",
    doi = "10.1103/PhysRevD.93.114034",
    journal = "Phys. Rev. D",
    volume = "93",
    number = "11",
    pages = "114034",
    year = "2016"
}

@article{Adams:2018pwt,
    author = "Adams, B. and others",
    title = "{Letter of Intent: A New QCD facility at the M2 beam line of the CERN SPS (COMPASS++/AMBER)}",
    eprint = "1808.00848",
    archivePrefix = "arXiv",
    primaryClass = "hep-ex",
    reportNumber = "CERN-SPSC-2019-003, SPSC-I-250",
    month = "8",
    year = "2018"
}

@article{Hedditch:2007ex,
    author = "Hedditch, J. N. and Kamleh, W. and Lasscock, B. G. and Leinweber, D. B. and Williams, A. G. and Zanotti, J. M.",
    title = "{Pseudoscalar and vector meson form-factors from lattice QCD}",
    eprint = "hep-lat/0703014",
    archivePrefix = "arXiv",
    reportNumber = "ADP-07-02-T642",
    doi = "10.1103/PhysRevD.75.094504",
    journal = "Phys. Rev. D",
    volume = "75",
    pages = "094504",
    year = "2007"
}

@article{Detmold:2017oqb,
    author = "Detmold, W. and Pefkou, D. and Shanahan, P. E.",
    title = "{Off-forward gluonic structure of vector mesons}",
    eprint = "1703.08220",
    archivePrefix = "arXiv",
    primaryClass = "hep-lat",
    reportNumber = "MIT-CTP-4894, MIT-CTP/4894",
    doi = "10.1103/PhysRevD.95.114515",
    journal = "Phys. Rev. D",
    volume = "95",
    number = "11",
    pages = "114515",
    year = "2017"
}

@article{Wang:2025hew,
    author = "Wang, Zhengli and Leinweber, Derek B. and Liu, Chuan and Liu, Liuming and Sun, Peng and Thomas, Anthony W. and Wu, Jia-jun and Xing, Hanyang and Yu, Kang",
    collaboration = "CLQCD",
    title = "{Spectral parameters of the {\ensuremath{\rho}} resonance from lattice QCD}",
    eprint = "2502.03700",
    archivePrefix = "arXiv",
    primaryClass = "hep-lat",
    reportNumber = "ADP-25-4/1266",
    doi = "10.1007/JHEP08(2025)064",
    journal = "JHEP",
    volume = "08",
    pages = "064",
    year = "2025"
}

@article{Bentz:2001vc,
    author = "Bentz, Wolfgang and Thomas, Anthony William",
    title = "{The Stability of nuclear matter in the Nambu-Jona-Lasinio model}",
    eprint = "nucl-th/0105022",
    archivePrefix = "arXiv",
    reportNumber = "ADP-01-09-T444",
    doi = "10.1016/S0375-9474(01)01119-8",
    journal = "Nucl. Phys. A",
    volume = "696",
    pages = "138--172",
    year = "2001"
}

@article{Whittenbury:2015ziz,
    author = "Whittenbury, D. L. and Matevosyan, H. H. and Thomas, A. W.",
    title = "{Hybrid stars using the quark-meson coupling and proper-time Nambu{\textendash}Jona-Lasinio models}",
    eprint = "1511.08561",
    archivePrefix = "arXiv",
    primaryClass = "nucl-th",
    doi = "10.1103/PhysRevC.93.035807",
    journal = "Phys. Rev. C",
    volume = "93",
    number = "3",
    pages = "035807",
    year = "2016"
}

@article{Hutauruk:2025wkn,
    author = "Hutauruk, Parada T. P.",
    title = "{Pseudoscalar Meson Parton Distributions Within Gauge-Invariant Nonlocal Chiral Quark Model}",
    eprint = "2505.06726",
    archivePrefix = "arXiv",
    primaryClass = "hep-ph",
    doi = "10.3390/sym17060971",
    journal = "Symmetry",
    volume = "17",
    number = "6",
    pages = "971",
    year = "2025"
}

@article{Son:2024uet,
    author = "Son, Hyeon-Dong and Hutauruk, Parada T. P.",
    title = "{Generalized parton distributions of the kaon and pion within the nonlocal chiral quark model}",
    eprint = "2411.18130",
    archivePrefix = "arXiv",
    primaryClass = "hep-ph",
    doi = "10.1103/PhysRevD.111.054007",
    journal = "Phys. Rev. D",
    volume = "111",
    number = "5",
    pages = "054007",
    year = "2025"
}

@article{Meissner:2026zos,
    author = "Mei{\ss}ner, Ulf-G. and Rusetsky, Akaki and Sakthivasan, Ajay S. and Schierholz, Gerrit and Wu, Jia-Jun",
    title = "{Form factors of the {\ensuremath{\rho}} meson from effective field theory and the lattice}",
    eprint = "2602.23044",
    archivePrefix = "arXiv",
    primaryClass = "hep-lat",
    reportNumber = "DESY-26-030",
    doi = "10.1007/JHEP06(2026)164",
    journal = "JHEP",
    volume = "06",
    pages = "164",
    year = "2026"
}

@article{Choi:2004ww,
    author = "Choi, Ho-Meoyng and Ji, Chueng-Ryong",
    title = "{Electromagnetic structure of the rho meson in the light front quark model}",
    eprint = "hep-ph/0402114",
    archivePrefix = "arXiv",
    doi = "10.1103/PhysRevD.70.053015",
    journal = "Phys. Rev. D",
    volume = "70",
    pages = "053015",
    year = "2004"
}

@article{Hecht:1997uj,
    author = "Hecht, M. B. and McKellar, B. H. J.",
    title = "{Dipole moments of the rho meson}",
    eprint = "hep-ph/9704326",
    archivePrefix = "arXiv",
    reportNumber = "UM-P-97-15",
    doi = "10.1103/PhysRevC.57.2638",
    journal = "Phys. Rev. C",
    volume = "57",
    pages = "2638--2647",
    year = "1998"
}

@article{Braguta:2004kx,
    author = "Braguta, V. V. and Onishchenko, A. I.",
    title = "{rho meson form-factors and QCD sum rules}",
    eprint = "hep-ph/0403258",
    archivePrefix = "arXiv",
    reportNumber = "WSU-HEP-0404",
    doi = "10.1103/PhysRevD.70.033001",
    journal = "Phys. Rev. D",
    volume = "70",
    pages = "033001",
    year = "2004"
}

@article{Aliev:2004uj,
    author = "Aliev, T. M. and Savci, M.",
    title = "{Electromagnetic form factors of the rho meson in light cone QCD sum rules}",
    eprint = "hep-ph/0405235",
    archivePrefix = "arXiv",
    reportNumber = "METU-PHYS-HEP-0405019",
    doi = "10.1103/PhysRevD.70.094007",
    journal = "Phys. Rev. D",
    volume = "70",
    pages = "094007",
    year = "2004"
}

@article{Jaus:2002sv,
    author = "Jaus, Wolfgang",
    title = "{Consistent treatment of spin 1 mesons in the light front quark model}",
    eprint = "hep-ph/0212098",
    archivePrefix = "arXiv",
    doi = "10.1103/PhysRevD.67.094010",
    journal = "Phys. Rev. D",
    volume = "67",
    pages = "094010",
    year = "2003"
}

@article{Carrillo-Serrano:2015uca,
    author = {Carrillo-Serrano, Manuel E. and Bentz, Wolfgang and Clo{\"e}t, Ian C. and Thomas, Anthony W.},
    title = "{$\rho$ meson form factors in a confining Nambu{\textendash}Jona-Lasinio model}",
    eprint = "1504.08119",
    archivePrefix = "arXiv",
    primaryClass = "nucl-th",
    reportNumber = "ADP-15-17-T919",
    doi = "10.1103/PhysRevC.92.015212",
    journal = "Phys. Rev. C",
    volume = "92",
    number = "1",
    pages = "015212",
    year = "2015"
}

@article{Hutauruk:2025bjd,
    author = "Hutauruk, Parada T. P. and Mart, Terry and Tsushima, Kazuo",
    title = "{Medium effects on the electromagnetic form factors of the {\ensuremath{\rho}} meson}",
    eprint = "2508.20501",
    archivePrefix = "arXiv",
    primaryClass = "nucl-th",
    reportNumber = "LFTC-25-08/102",
    doi = "10.1103/95yl-nwds",
    journal = "Phys. Rev. D",
    volume = "112",
    number = "11",
    pages = "114030",
    year = "2025"
}

@article{Cloet:2014rja,
    author = {Clo{\"e}t, Ian C. and Bentz, Wolfgang and Thomas, Anthony W.},
    title = "{Role of diquark correlations and the pion cloud in nucleon elastic form factors}",
    eprint = "1405.5542",
    archivePrefix = "arXiv",
    primaryClass = "nucl-th",
    doi = "10.1103/PhysRevC.90.045202",
    journal = "Phys. Rev. C",
    volume = "90",
    pages = "045202",
    year = "2014"
}

@article{Roberts:2011wy,
    author = "Roberts, H. L. L. and Bashir, A. and Gutierrez-Guerrero, L. X. and Roberts, C. D. and Wilson, D. J.",
    title = "{pi- and rho-mesons, and their diquark partners, from a contact interaction}",
    eprint = "1102.4376",
    archivePrefix = "arXiv",
    primaryClass = "nucl-th",
    doi = "10.1103/PhysRevC.83.065206",
    journal = "Phys. Rev. C",
    volume = "83",
    pages = "065206",
    year = "2011"
}

@article{Gutierrez-Guerrero:2026rsb,
    author = "Guti{\'e}rrez-Guerrero, Laura Xiomara and Hern{\'a}ndez-Pinto, Roger Jos{\'e}",
    title = "{Symmetry- Preserving Contact Interaction Approaches: An Overview of Meson and Diquark Form Factors}",
    eprint = "2604.15122",
    archivePrefix = "arXiv",
    primaryClass = "hep-ph",
    doi = "10.3390/particles9020045",
    journal = "Particles",
    volume = "9",
    number = "2",
    pages = "45",
    year = "2026"
}

@article{Luan:2015goa,
    author = "Luan, Yi-Long and Chen, Xiao-Lin and Deng, Wei-Zhen",
    title = "{Meson electro-magnetic form factors in an extended Nambu{\textendash}Jona-Lasinio model including heavy quark flavors}",
    eprint = "1504.03799",
    archivePrefix = "arXiv",
    primaryClass = "hep-ph",
    doi = "10.1088/1674-1137/39/11/113103",
    journal = "Chin. Phys. C",
    volume = "39",
    number = "11",
    pages = "113103",
    year = "2015"
}

@article{Luschevskaya:2026kxx,
    author = "Luschevskaya, E. V. and Teryaev, O. V. and Dorenskaya, E. A. and Alimagomedova, S. Y. and Khaidukov, Z. V.",
    title = "{Magnetic Moment of K*$^{0}$ Mesons in SU(3) Lattice Gauge Theory}",
    doi = "10.1134/S0021364025608437",
    journal = "JETP Lett.",
    volume = "123",
    number = "10",
    pages = "653--660",
    year = "2026"
}

@article{Lee:2008qf,
    author = "Lee, Frank X. and Moerschbacher, Scott and Wilcox, Walter",
    title = "{Magnetic moments of vector, axial, and tensor mesons in lattice QCD}",
    eprint = "0807.4150",
    archivePrefix = "arXiv",
    primaryClass = "hep-lat",
    doi = "10.1103/PhysRevD.78.094502",
    journal = "Phys. Rev. D",
    volume = "78",
    pages = "094502",
    year = "2008"
}

@article{Badalian:2012ft,
    author = "Badalian, A. M. and Simonov, Yu. A.",
    title = "{Magnetic moments of mesons}",
    eprint = "1211.4349",
    archivePrefix = "arXiv",
    primaryClass = "hep-ph",
    doi = "10.1103/PhysRevD.87.074012",
    journal = "Phys. Rev. D",
    volume = "87",
    number = "7",
    pages = "074012",
    year = "2013"
}
%----------------------------------------------------------------------------------------
\end{document}